\documentclass[12pt]{article}  
  
\usepackage{epsfig,epsf}  
\usepackage{graphics}  
\usepackage{cite}  
\usepackage{geometry}
\usepackage{xcolor} 

\usepackage{amsmath}  
\usepackage{amsthm}  
\usepackage{amsfonts}  
\usepackage{amssymb} 
\usepackage{bm} 

\usepackage{slashed}
\usepackage{braket}

\usepackage{hyperref}
\usepackage{longtable}
\usepackage{caption}

\numberwithin{equation}{section}
\newcommand{\Lbar}{\overline{\Lambda}}

\begin{document}
\allowdisplaybreaks
\thispagestyle{empty}  
  
\begin{flushright}  
{\small  
IPPP/19/20 \\[0.1cm]
\today
}  
\end{flushright}  
  
\vskip1.5cm  
\begin{center}  
\textbf{\Large\boldmath $B_s$ mixing observables and $|V_{td}/V_{ts}|$ from sum rules}  
\\  
\vspace{1.5cm}  
{\sc Daniel~King}$^{(a)}$, {\sc Alexander~Lenz}$^{(a)}$ and {\sc Thomas~Rauh}$^{(a,b)}$\\[0.5cm]  
\vspace*{0.5cm} {\it 
$(a)$ IPPP, Department of Physics,
University of Durham,\\
DH1 3LE, United Kingdom\\[0.2cm]
$(b)$ Albert Einstein Center for Fundamental Physics,\\
Institute for Theoretical Physics, University of Bern,\\
Sidlerstrasse 5, CH-3012 Bern, Switzerland }
  
\def\thefootnote{\arabic{footnote}}  
\setcounter{footnote}{0}  
  
\vskip2.5cm  
\textbf{Abstract}\\  
\vspace{1\baselineskip}  
\parbox{0.9\textwidth}{
We consider the effects of a non-vanishing strange-quark mass in the 
determination of the full basis of dimension six matrix elements for 
$B_{s}$ mixing, in particular we get for the ratio of the $V-A$ Bag parameter 
in the $B_s$ and $B_d$ system: $\overline{B}^s_{Q_1} / \overline{B}^d_{Q_1} = 0.987^{+0.007}_{-0.009}$.
Combining these results with the most recent lattice values for the ratio of decay constants
$f_{B_s} / f_{B_d}$ we obtain the most precise determination of the ratio 
$\xi = f_{B_s} \sqrt{\overline{B}^s_{Q_1}}/ f_{B_d} \sqrt{\overline{B}^d_{Q_1}} 
= 1.2014^{+0.0065}_{-0.0072}$ in agreement with recent lattice determinations. 
We find $\Delta M_s=(18.5_{-1.5}^{+1.2})\text{ps}^{-1}$ and $\Delta M_d=(0.547_{-0.046}^{+0.035})\text{ps}^{-1}$ 
to be consistent with experiments at below one sigma. Assuming the validity of the SM, 
our calculation can be used to directly determine the ratio of CKM elements
$|V_{td} / V_{ts} | = 0.2045^{+0.0012}_{-0.0013}$, which is compatible with the results 
from the CKM fitting groups, but again more precise.}  
  
\end{center}  
  
  
\newpage  
\setcounter{page}{1}


\section{Introduction\label{sec:intro}}

Mixing of $B_s$ mesons is experimentally well studied \cite{Artuso:2015swg} and the mass
difference $\Delta M_s = 2 |M_{12}^s|$ is known with a high precision \cite{Amhis:2016xyh} (based on the
individual measurements
\cite{Abulencia:2006ze,Aaij:2011qx,Aaij:2013mpa,Aaij:2013gja,Aaij:2014zsa}):
\begin{equation}
  \Delta M_s^{\rm Exp.} = (17.757 \pm 0.021) \,  \mbox{ps}^{-1} \, .
\end{equation}
The  corresponding theory expression for  $M_{12}^s$ reads
\begin{eqnarray}
M_{12}^s & = & \frac{G_F^2}{12 \pi^2} \lambda_t^2 M_W^2 S_0(x_t) \hat{\eta }_B B f_{B_s}^2  M_{B_s} \, ,
\label{M12}
\end{eqnarray}
with the CKM element $\lambda_t = V_{ts}^* V_{tb}$ and the
Inami-Lim function  $S_0$ \cite{Inami:1980fz} describing the result of the 1-loop box diagrams
in the standard model (SM). Perturbative 2-loop QCD corrections are compressed in the factor
$\hat{\eta }_B $ \cite{Buras:1990fn}.
Since this observable is loop-suppressed in the SM, it is expected to be very sensitive to BSM effects.
The bag parameter $B\equiv\overline{B}_{Q_1}^s$ and the decay constant $f_{B_s}$
quantify the hadronic contribution to $B$-mixing; the uncertainties of their numerical values make
up the biggest uncertainty by far in the SM prediction of the mass difference. These parameters have
been determined
by lattice simulations \cite{Dalgic:2006gp,Carrasco:2013zta,Bazavov:2016nty} and for the case of $B_d$ mesons 
with HQET sum rules~\cite{Grozin:2016uqy,Grozin:2017uto,Kirk:2017juj,Grozin:2018wtg}. There is also a recent 
lattice determination of the SU(3) breaking ratios~\cite{Boyle:2018knm}. 
\\
Taking the most recent lattice average from the Flavour Lattice Averaging Group (FLAG) \cite{Aoki:2019cca},
which is more or less equivalent to the result in \cite{Bazavov:2016nty}, one gets \cite{DiLuzio:2017fdq}
a SM prediction for the mass difference, which is larger than the measurement:
\begin{equation}
  \Delta M_s^{\rm SM, 2017} = (20.01 \pm 1.25) \,  \mbox{ps}^{-1} \, .
\label{DeltaMsSM1}
\end{equation}
Such a value has dramatic consequences for some of the BSM models that are  currently investigated in order
to explain the flavour anomalies. In particular the parameter space of certain $Z'$ models is almost completely
excluded \cite{DiLuzio:2017fdq}.
\\
In this work we  extend the analysis of \cite{Kirk:2017juj} with effects of a finite 
strange-quark mass, thus getting for the first time a HQET sum rule prediction for the mixing Bag
parameter of $B_s$ mesons. Lattice simulations typically achieve a much higher precision than sum rule
calculations, but in our case a sum rule for $B-1$ can be written down. Since the value of the
Bag parameter $B$ is close to 1, even a moderate precision of the sum rule of the order of 20 $\%$
for $B-1$, turns into a precision of the order of $2 \%$ for the whole Bag parameter, which is highly
competitive. Thus our determination constitutes an independent cross-check of the large lattice value found in
\cite{Bazavov:2016nty}.
In combination with a precise lattice determination of the decay constant $f_{B_s}$ our result for the Bag
parameter can also be used for a direct determination of $|V_{ts}^*V_{tb}|$ from the measured mass
difference $\Delta M_s^{\rm Exp.}$. Taking instead a ratio of the mass differences in the $B_d$ and the $B_s$
system one can get a clean handle on  $|V_{td}/V_{ts}|$. Taking further a ratio of $\Delta M_s$ and the rare branching ratio
$Br (B_s \to \mu^+ \mu^-)$ the decay constant and the CKM dependence cancel and the Bag parameter will be
the only relevant input parameter. 
\\
Our paper is organised as follows: in Section \ref{sec:sum_rule} we set up the sum rule for the Bag parameter and
determine the $m_s$ corrections, in Section \ref{sec:results} we present a numerical study of the sum rules
and we perform a phenomenological analysis. Finally, we conclude in Section \ref{sec:conclusion}.


\section{Sum rules in HQET\label{sec:sum_rule}}


\subsection{Operator basis and definition of bag parameters\label{sec:operators}}

In this work we use the full dimension-six $\Delta B=2$ operator basis required for a calculation 
of $\Delta M_s$ in the SM\footnote{The operator $Q_1$ corresponds to the SM contribution to $\Delta M_s$.} 
and BSM theories and for a SM prediction of $\Delta\Gamma_s$. The QCD operators involved are
\begin{eqnarray}
 Q_1 & = & \bar{b}_i\gamma_\mu(1-\gamma^5)s_i\,\,\bar{b}_j\gamma^\mu(1-\gamma^5)s_j, \nonumber\\
 Q_2 & = & \bar{b}_i(1-\gamma^5)s_i\,\,\bar{b}_j(1-\gamma^5)s_j, \hspace{1cm} Q_3 = \bar{b}_i(1-\gamma^5)s_j\,\,\bar{b}_j(1-\gamma^5)s_i,\nonumber\\
 Q_4 & = & \bar{b}_i(1-\gamma^5)s_i\,\,\bar{b}_j(1+\gamma^5)s_j, \hspace{1cm} Q_5 = \bar{b}_i(1-\gamma^5)s_j\,\,\bar{b}_j(1+\gamma^5)s_i. 
 \label{eq:QCD_operators}
\end{eqnarray} 
while our HQET basis is defined as
\begin{eqnarray}
 \tilde{Q}_1 & = & \bar{h}_i^{\{(+)}\gamma_\mu(1-\gamma^5)s_i\,\,\bar{h}_j^{(-)\}}\gamma^\mu(1-\gamma^5)s_j, \hspace{0.5cm}
 \tilde{Q}_2 = \bar{h}_i^{\{(+)}(1-\gamma^5)s_i\,\,\bar{h}_j^{(-)\}}(1-\gamma^5)s_j,\nonumber\\
 \tilde{Q}_4 & = & \bar{h}_i^{\{(+)}(1-\gamma^5)s_i\,\,\bar{h}_j^{(-)\}}(1+\gamma^5)s_j, \hspace{1.35cm} 
 \tilde{Q}_5 = \bar{h}_i^{\{(+)}(1-\gamma^5)s_j\,\,\bar{h}_j^{(-)\}}(1+\gamma^5)s_i,
 \nonumber
 \\
 \label{eq:HQET_operators}
\end{eqnarray}
where $h^{(+/-)}(x)$ is the HQET bottom/anti-bottom field and we use the notation
\begin{equation}
 \bar{h}^{\{(+)}\Gamma_A s\,\,\bar{h}^{(-)\}}\Gamma_B s = \bar{h}^{(+)}\Gamma_A s\,\,\bar{h}^{(-)}\Gamma_B s + \bar{h}^{(-)}\Gamma_A s\,\,\bar{h}^{(+)}\Gamma_B s.
\end{equation}
The matching condition is given by
\begin{equation}
    \braket{Q_i}(\mu)=\sum C_{Q_i \tilde{Q}_j}\braket{\tilde{Q}_j}+\mathcal{O}(1/m_b),
    \label{matching}
\end{equation}
for which the NLO HQET-QCD matching coefficients $C_{Q\tilde{Q}}$ were presented in \cite{Kirk:2017juj}. 
We also use the same basis of evanescent operators. As mentioned in \cite{Kirk:2017juj}, the HQET 
evanescent operators are defined up to 3 constants $a_i$ with $i={1,2,3}$ in order to gauge the scheme 
dependence. We also note that in all of the following we work within the NDR scheme in dimensional 
regularisation with $d=4-2\epsilon$.
\\
The QCD bag parameters $B_Q^s$ are defined through \cite{Gabbiani:1996hi}
\begin{equation}
  \braket{Q(\mu)} = A_Q\, f_{B_s}^2M_{B_s}^2\, B_Q^s(\mu) = \overline{A}_Q(\mu)\, f_{B_s}^2M_{B_s}^2\, \overline{B}_Q^s(\mu),
  \label{eq:Bags_QCD}
\end{equation}
with the coefficients $A_Q$ given by
\begin{equation}
 \begin{array}{ll}
  A_{Q_1} = 2+\frac{2}{N_c}, & \\
  A_{Q_2} = \frac{M_{B_s}^2}{(m_b+m_s)^2}\left(-2+\frac{1}{N_c}\right), \hspace{1cm}& A_{Q_3} = \frac{M_{B_s}^2}{(m_b+m_s)^2}\left(1-\frac{2}{N_c}\right),\\
  A_{Q_4} = \frac{2M_{B_s}^2}{(m_b+m_s)^2}+\frac{1}{N_c}, & A_{Q_5} = 1+\frac{2M_{B_s}^2}{N_c(m_b+m_s)^2}, 
 \end{array}
 \label{eq:AQi_QCD}
\end{equation}
where $M_{B_s}$ denotes the $B_s$ meson mass, $m_q$ corresponds to quark pole masses and the $B_s$ meson decay constant $f_{B_s}$ is defined by
\begin{equation}
 \braket{0|\bar{b}\gamma^\mu\gamma^5 s|B_s(p)} = -if_{B_s}p^\mu. 
\end{equation}
The barred terms in the far right expression of (\ref{eq:Bags_QCD}) indicate that the quark masses used there are in the $\overline{\text{MS}}$ scheme. For the reasons discussed in \cite{Kirk:2017juj} we prefer to use the pole masses for our analysis and then convert to this form at the end. Similarly, the HQET bag parameters are defined through
\begin{equation}
 \hm{\langle}\hspace{-0.2cm}\hm{\langle}\tilde{Q}(\mu)\hm{\rangle}\hspace{-0.2cm}\hm{\rangle} = 
 A_{\tilde{Q}}\, F_s^2(\mu)\, B_{\tilde{Q}}^s(\mu), 
 \label{eq:Bags_HQET}
\end{equation}
with the coefficients $A_{\tilde{Q}}$ given by
\begin{equation}
  A_{\tilde{Q}_1} = 2+\frac{2}{N_c},\hspace{0.4cm} A_{\tilde{Q}_2} = -2+\frac{1}{N_c},\hspace{0.4cm}
  A_{\tilde{Q}_4} = 2+\frac{1}{N_c},\hspace{0.4cm} A_{\tilde{Q}_5} = 1+\frac{2}{N_c},
 \label{eq:AQi_HQET}
\end{equation}
and where the matrix elements are taken between non-relativistically normalised states 
$\hm{\langle}\hspace{-0.2cm}\hm{\langle}\tilde{Q}(\mu)\hm{\rangle}\hspace{-0.2cm}\hm{\rangle}\equiv \braket{\overline{\bold{B}}_s|\tilde{Q}(\mu)|\bold{B}_s}$ with
\begin{equation}
    |B_s(p)\rangle =\sqrt{2 M_{B_s}}|\bold{B}_s(v)\rangle +\mathcal{O}(1/m_b).
 \label{eq:B state}
\end{equation}
The HQET decay constant $F_s(\mu)$, appearing in (\ref{eq:Bags_HQET}) is defined by
\begin{equation}
 \braket{0|\bar{h}^{(-)}\gamma^\mu\gamma^5 s|\mathbf{B}_s(v)} = -iF_s(\mu)v^\mu, 
\end{equation}
which is then related to the QCD decay constant $f_{B_s}$ through
\begin{equation}
 f_{B_s} = \sqrt{\frac{2}{M_{B_s}}}C(\mu)F_s(\mu)+\mathcal{O}\left(1/m_b\right),
 \label{eq:fB_F}
\end{equation}
with~\cite{Eichten:1989zv} 
\begin{equation}
 C(\mu) = 1-2C_F\frac{\alpha_s(\mu)}{4\pi}+\mathcal{O}(\alpha_s^2).
 \label{eq:C(mu)}
\end{equation}
From our sum rule analysis we determine the HQET bag parameters $B_{\tilde{Q}}^s$. Using (\ref{matching}), (\ref{eq:Bags_QCD}), (\ref{eq:Bags_HQET}), and (\ref{eq:fB_F}) we arrive at the relation
\begin{equation}
 B_{Q_i}^s(\mu) = \sum\limits_j \frac{A_{\tilde{Q}_j}}{A_{Q_i}}\,\frac{C_{Q_i\tilde{Q}_j}(\mu)}{C^2(\mu)}\,B_{\tilde{Q}_j}^s(\mu) + \mathcal{O}(1/m_b),
 \label{eq:B_QCD_HQET_matching}
\end{equation}
which allows us to then match the values of $B_{\tilde{Q}}^s$ onto their QCD counterparts.


\subsection{\boldmath Finite $m_s$ effects in the HQET decay constant\label{sec:sum_rule_decay_const}}

To illustrate our strategy for the treatment of finite $m_s$ effects we 
first consider the Borel sum rule for the HQET decay constant $F_s$ which 
has been derived in~\cite{Broadhurst:1991fc,Bagan:1991sg,Neubert:1991sp}. 
In the $B_s$ system it takes the form  
\begin{equation}
 F_s^2(\mu_\rho)e^{-\frac{\Lbar+m_s}{t}}=\int\limits_0^{\omega_c}d\omega\, e^{-\frac{\omega}{t}}\rho_\Pi(\omega) \, ,
 \label{eq:SR_F2}
\end{equation}
where $\rho_\Pi$ is the discontinuity of the two-point correlator 
\begin{equation}
 \Pi(\omega) = i\,\int d^dx e^{ipx}\braket{0|\text{T}\left[\tilde{j}_+^\dagger(0)\tilde{j}_+(x)\right]|0} \, ,
 \label{eq:Pi}
\end{equation}
with $\omega=p\cdot v$ and the interpolating current $\tilde{j}_{+} = \bar{s}\gamma^5 h^{(+)}$.
The leading perturbative part of the discontinuity is given by 
\begin{equation}
 \rho_\Pi^\text{pert}(\omega) = \frac{N_c}{2\pi^2}\left[(\omega+m_s)\sqrt{\omega^2-m_s^2}\,\theta(\omega-m_s) + \mathcal{O}(\alpha_s)\right] \, .
 \label{eq:rho_Pi_ms}
\end{equation}
In the remainder of this subsection we consider the finite-energy (FESR) 
version of the sum rule~\eqref{eq:SR_F2} which is given by the limit 
$t\to\infty$ to be able to present compact analytic results. We obtain 
\begin{eqnarray}
 F_s^2(\mu_\rho)|_\text{FESR} & = & \frac{N_c}{6\pi^2}\Bigg[
      \left(\omega_c-\frac{m_s}{2}\right)(\omega_c+2m_s)\sqrt{\omega_c^2-m_s^2} \nonumber\\
 & &  + \frac{3m_s^3}{2}\ln\left(\frac{m_s}{\omega_c+\sqrt{\omega_c^2-m_s^2}}\right) 
      + \mathcal{O}(\alpha_s) + [\text{condensates}]\Bigg] \nonumber\\
& = & \frac{N_c\omega_c^3}{6\pi^2}\left[1+\frac{3m_s}{2\omega_c}-\frac{3 m_s^2}{2 \omega_c^2} - 
      \frac{3 m_s^3}{4 \omega_c^3}\left(1 - \ln\frac{m_s^2}{4 \omega_c^2}\right) + \dots \right] \, .
\label{eq:F2_FESR}
\end{eqnarray}
In the last step we have expanded the result in the small ratio 
$m_s/\omega_c\sim 0.1$. The appearance of a $m_s^3\ln(m_s)$ term in the 
expansion indicates that energies $\omega$ of the order $m_s$ contribute at 
order $m_s^3$. These logarithms can be absorbed into the quark condensate 
\cite{Broadhurst:1991fc,Generalis:1989hf}. In the following we show how the 
terms up to order $m_s^2$ can be determined without knowing the full 
$m_s$ dependence of the discontinuity~\eqref{eq:rho_Pi_ms}. This 
will be essential for the determination of the $m_s$ effects in 
the Bag parameters where the calculation of the full $m_s$ 
dependence is very challenging (3 loops and 3 scales). We 
first split the integration at an arbitrary scale $\nu$ with 
$m_s\ll\nu\ll\omega_c$. Above $\nu$ we may expand the integrand 
in $m_s/\omega$, yielding the identity 
\begin{equation}
 \mathcal{T}_{\frac{m_s}{\omega_c}}[F_s^2(\mu_\rho)]e^{-\frac{\Lbar+m_s}{t}} = 
 \mathcal{T}_{\{\frac{m_s}{\omega_c},\frac{m_s}{\nu},\frac{\nu}{\omega_c}\}}\left[\,\, \int\limits_{m_s}^{\nu}d\omega\, e^{-\frac{\omega}{t}}\rho_\Pi(\omega) + 
 \int\limits_\nu^{\omega_c}d\omega\, e^{-\frac{\omega}{t}}\mathcal{T}_{\frac{m_s}{\omega}}[\rho_\Pi(\omega)] \,\right]\,,
 \label{eq:SR_F2_exp}
\end{equation}
where $\mathcal{T}_{x}[\dots]$ indicates that the expression in 
square brackets must be Taylor expanded in $x$. 
The dependence on the scale $\nu$ has to cancel in the expanded 
result. We can therefore take the limit $\nu\to m_s$ \emph{after} 
expanding the result according to the scaling $m_s\ll\nu\ll\omega_c$. 
We note that the contribution from the integration of the full 
integrand between $m_s$ and $\nu$ does not vanish for $\nu\to m_s$, 
because the limit has to be taken after the expansion in $m_s$ and 
the two operations do not commute. It is however clear from 
dimensional analysis that this contribution must be polynomial  
in $m_s$ starting at $m_s^3$ since the exponential can be Taylor 
expanded. This demonstrates that it is sufficient to compute the 
discontinuity~\eqref{eq:rho_Pi_ms} as an expansion in $m_s/\omega$ 
if we restrict the analysis to the linear and quadratic terms which 
is clearly sufficient due to the small expansion parameter. 
In the FESR limit considered above we find\footnote{Here the limit 
$\nu\to m_s$ and the Taylor expansion commute, because the integrand 
is polynomial in $m_s$.} 
\begin{equation}
 \mathcal{T}_{\frac{m_s}{\omega_c}}\left[\,\, \int\limits_{m_s}^{\omega_c}d\omega\, \mathcal{T}_{\frac{m_s}{\omega}}[\rho_\Pi(\omega)] \,\right] = 
 \frac{N_c\omega_c^3}{6\pi^2}\left[1+\frac{3m_s}{2\omega_c}-\frac{3 m_s^2}{2 \omega_c^2} - 
      \frac{m_s^3}{\omega_c^3}\left(1 - \frac{3}{4}\ln\frac{m_s^2}{\omega_c^2}\right) + \dots \right] \, .
\label{eq:F2_FESR_exp}
\end{equation}
The difference between~\eqref{eq:F2_FESR} and~\eqref{eq:F2_FESR_exp} 
is indeed of order $m_s^3$ and is compensated by the contribution 
from the first term on the right-hand side of~\eqref{eq:SR_F2_exp}. 

\begin{figure}
 \begin{center}
    \includegraphics[width=0.2\textwidth]{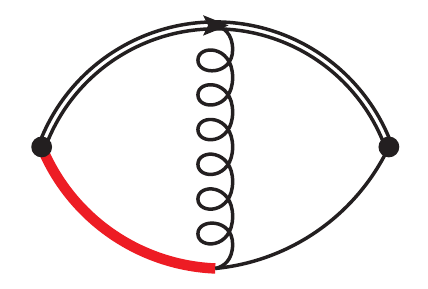}
  \caption{\label{fig:2pt_soft}
  Sample diagram involving a soft light-quark propagator (red thick line).}
 \end{center}
\end{figure}

At NLO we therefore only compute the expanded result by using the 
method of regions~\cite{Beneke:1997zp,Jantzen:2011nz}. The light 
degrees of freedom can be either hard with momentum $k\sim\omega$ 
or soft with momentum $k\sim m_s$ whereas the heavy quark field is 
always hard. Up to and including the order $m_s^2$ there are 
however only contributions from diagrams where all lines are hard. 
An example diagram involving a soft line is shown in Figure~\ref{fig:2pt_soft}. 
The integral measure scales as $m_s^4$ and the soft light-quark 
propagator scales as $m_s^{-1}$, yielding an overall scaling of 
$m_s^3$. Diagrams where only the gluon is soft are scaleless and 
vanish in dimensional regularization. Contributions where both 
loop momenta are soft are of the order $m_s^4$. 
Therefore, we only need to consider the fully hard momentum region 
where the integrand can be naively Taylor expanded in $m_s$. 
We obtain 
\begin{eqnarray}
 \rho_\Pi(\omega) & \equiv & \frac{\Pi(\omega+i0)-\Pi(\omega-i0)}{2\pi i} \\
                  &    =   & \frac{N_c\omega^2}{2\pi^2}\,\theta(\omega-m_s)\Bigg\{1 + \frac{m_s}{\omega} - \frac12\,\left(\frac{m_s}{\omega}\right)^2 + \dots \nonumber\\
                  & & + \frac{\alpha_sC_F}{4\pi}\,\Bigg[
                  17 + \frac{4\pi^2}{3} + 3\ln\frac{\mu_\rho^2}{4\omega^2} + 
                  \left(20 + \frac{4\pi^2}{3} + 6\ln\frac{\mu_\rho^2}{4\omega^2} - 3\ln\frac{\mu_\rho^2}{m_s^2}\right)\,\frac{m_s}{\omega} \nonumber\\
                  & & + \left(1 - \frac92 \ln\frac{\mu_\rho^2}{4\omega^2} + 3\ln\frac{\mu_\rho^2}{m_s^2}\right)\,\left(\frac{m_s}{\omega}\right)^2 
                  + \dots\Bigg] + \mathcal{O}(\alpha_s^2)\Bigg\} + \text{[condensates]},\nonumber
 \label{eq:rho_Pi_ms_exp}
\end{eqnarray}
in agreement with \cite{Broadhurst:1991fc}.


\subsection{\boldmath Finite $m_s$ effects in the Bag parameters\label{sec:sum_rule_bags}}

The sum rule for the Bag parameters is based on the three-point correlator 
\begin{equation}
 K_{\tilde{Q}}(\omega_1,\omega_2) = \int d^dx_1 d^dx_2 e^{ip_1\cdot x_1-ip_2\cdot x_2}\braket{0|\text{T}\left[\tilde{j}_{+}(x_2)\tilde{Q}(0)\tilde{j}_{-}(x_1)\right]|0},
 \label{eq:DefK}
\end{equation}
where $\omega_{1,2} = p_{1,2}\cdot v$ and the interpolating currents for the 
$\overline{B}_s$ and $B_s$ mesons read 
\begin{equation}
 \tilde{j}_{+} = \bar{s}\gamma^5 h^{(+)}, \hspace{1cm} \tilde{j}_{-} = \bar{s}\gamma^5 h^{(-)}.
\end{equation}
The accuracy of the sum rule approach crucially depends on the observation 
that the contributions to the correlator can be split into factorizable and 
non-factorizable ones, examples of which are given in Figure~\ref{eq:DefK}. 
\begin{figure}
 \begin{center}
    \includegraphics[width=0.75\textwidth]{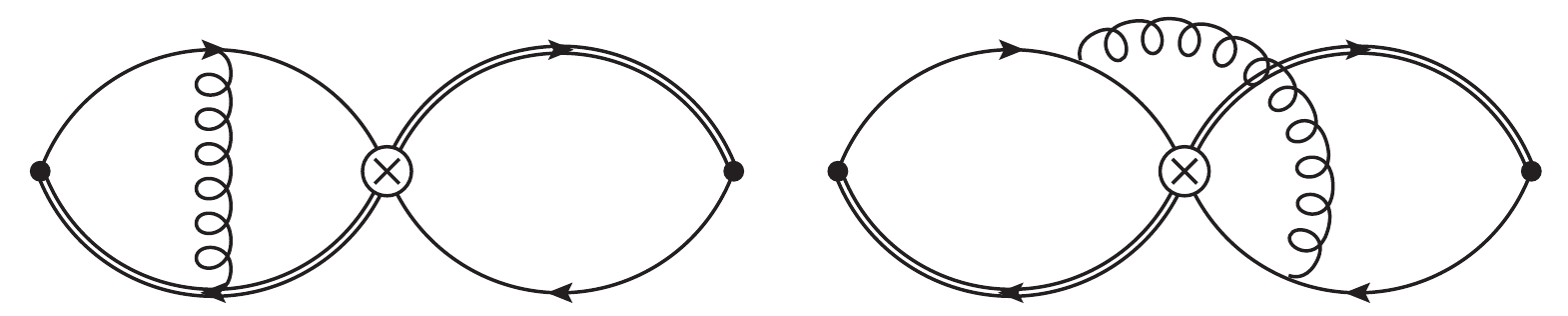}
  \caption{\label{fig:Correlator}
  Examples for factorizing (left) and non-factorizing (right) contributions 
  to the three-point correlator~\eqref{eq:DefK} at NLO in $\alpha_s$.}
 \end{center}
\end{figure}
The full set of factorizable contributions amounts to $B_{\tilde{Q}}^s = 1$ 
which allows us to formulate a sum rule for the deviation 
$\Delta B_{\tilde{Q}}^s = B_{\tilde{Q}}^s - 1$ based only on the non-factorizable 
contributions~\cite{Chetyrkin:1985vj,Korner:2003zk,Grozin:2016uqy,Kirk:2017juj}
\begin{eqnarray}
\label{eq:B_OneInt}
\Delta B_{\tilde{Q}_i}^s(\mu_\rho) & = & \frac{1}{A_{\tilde{Q}_i}F_s(\mu_\rho)^4}
\int\limits_0^{\omega_c}d\omega_1d\omega_2 e^{\frac{\Lbar+m_s-\omega_1}{t_1}+\frac{\Lbar+m_s-\omega_2}{t_2}}\Delta\rho_{\tilde{Q}_i}(\omega_1,\omega_2)\\
& = & \frac{1}{A_{\tilde{Q}_i}}\frac{\int\limits_0^{\omega_c}d\omega_1d\omega_2 e^{-\frac{\omega_1}{t_1}-\frac{\omega_2}{t_2}}\Delta\rho_{\tilde{Q}_i}(\omega_1,\omega_2)
      }{\left(\int\limits_0^{\omega_c}d\omega_1e^{-\frac{\omega_1}{t_1}}\rho_\Pi(\omega_1)\right)\left(\int\limits_0^{\omega_c}d\omega_2e^{-\frac{\omega_2}{t_2}}\rho_\Pi(\omega_2)\right)}.
\label{eq:B_ThreeInt}
\end{eqnarray}
where the second equation makes use of \eqref{eq:SR_F2}. The quantity 
$\Delta\rho_{\tilde{Q}_i}$ is the non-factorizable part of the double 
discontinuity 
\begin{equation}
 \rho_{\tilde{Q}_i}(\omega_1,\omega_2) = A_{\tilde{Q}_i}\rho_\Pi(\omega_1)\rho_\Pi(\omega_2)+\Delta\rho_{\tilde{Q}_i}\,.
 \label{eq:rho_decomposition}
\end{equation}
In~\cite{Kirk:2017juj} we derived a simple analytical result for the HQET 
bag parameters by comparing~\eqref{eq:B_OneInt} to the square of the sum 
rule for the decay constant~\eqref{eq:SR_F2} with an appropriately chosen 
weight function 
\begin{equation}
 w_{\tilde{Q}_i}(\omega_1,\omega_2) = 
    \frac{\Delta\rho_{\tilde{Q}_i}^\text{pert}(\omega_1,\omega_2)}{\rho_\Pi^\text{pert}(\omega_1)\rho_\Pi^\text{pert}(\omega_2)}\,.
 \label{eq:weight_function}
\end{equation}
The generalization of this approach to the $m_s$ corrections is straightforward. 
Expanding the double discontinuity in $m_s$, we obtain 
\begin{align}
 \Delta\rho_{\tilde{Q}_i}^\text{pert}(\omega_1,\omega_2) \equiv \,& \frac{N_c C_F}{4} \frac{\omega_1^2\omega_2^2}{\pi^4}\frac{\alpha_s}{4\pi}
 \Bigg[r_{\tilde{Q}_i}^{(0)}(x,L_\omega) + \left(\frac{m_s}{\omega_1} + \frac{m_s}{\omega_2}\right)r_{\tilde{Q}_i}^{(1)}(x,L_\omega) \nonumber\\
 & + \left(\frac{m_s^2}{\omega_1^2} + \frac{m_s^2}{\omega_2^2}\right)r_{\tilde{Q}_i}^{(2)}(x,L_\omega) 
 + \dots \Bigg]\,\theta(\omega_1-m_s)\theta(\omega_2-m_s),
 \label{eq:r_def}
\end{align}
where $x=\omega_2/\omega_1$ and $L_\omega=\ln(\mu_\rho^2/(4\omega_1\omega_2))$. 
With this parametrization, the symmetry of the three-point correlator 
under exchange of $\omega_1$ and $\omega_2$ manifests as a symmetry under 
$x\leftrightarrow1/x$ of the $r_{\tilde{Q}_i}^{(j)}$. The result for the 
deviation of the Bag parameters from the VSA reads 
\begin{align}
& \Delta B_{\tilde{Q}_i}^{s,\text{pert}}(\mu_\rho) = \frac{w_{\tilde{Q}_i}(\Lbar+m_s,\Lbar+m_s)}{A_{\tilde{Q}_i}} = \nonumber\\
& \frac{C_F}{N_c A_{\tilde{Q}_i}} \, \frac{\alpha_s(\mu_\rho)}{4\pi} \, 
    \Bigg\{r_{\tilde{Q}_i}^{(0)}\left(1,L_{\Lbar+m_s}\right) + 
    \frac{2m_s}{\Lbar+m_s}\left[r_{\tilde{Q}_i}^{(1)}\left(1,L_{\Lbar+m_s}\right) - r_{\tilde{Q}_i}^{(0)}\left(1,L_{\Lbar+m_s}\right)\right] \nonumber\\
& + \frac{2m_s^2}{(\Lbar+m_s)^2} \left[r_{\tilde{Q}_i}^{(2)}\left(1,L_{\Lbar+m_s}\right) 
      -2r_{\tilde{Q}_i}^{(1)}\left(1,L_{\Lbar+m_s}\right) + 2r_{\tilde{Q}_i}^{(0)}\left(1,L_{\Lbar+m_s}\right)\right] 
    + \dots \Bigg\}, 
 \label{eq:B_NoInt}
\end{align}
where $L_{\Lbar+m_s} = \ln(\mu_\rho^2/(4(\Lbar+m_s)^2))$. We find that 
the result only depends on the value of the double discontinuity at 
$\omega_1=\omega_2=\Lbar+m_s$. Thus, the knowledge of the $m_s$-expanded 
double discontinuity is sufficient to determine the $m_s$ effects for the 
Bag parameters in $B_s$ mixing. However, the use of this weight function 
approach relies on the expanded version of the sum rule~\eqref{eq:SR_F2} 
for the decay constant. As discussed in the previous subsection, this 
approach gives an incorrect result at the order $m_s^3$ and the result 
\eqref{eq:B_NoInt} is therefore limited to the quadratic order in $m_s$.


\subsection{Non-zero $m_s$ corrections to the non-factorizable part\label{sec:sum_rule_nonfac}}

\begin{figure}
 \begin{center}
    \includegraphics[width=0.75\textwidth]{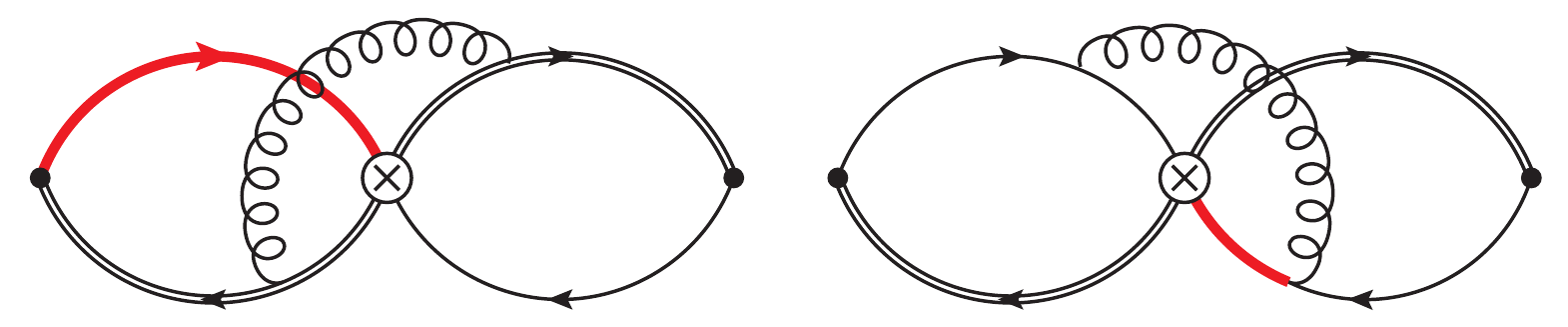}
  \caption{\label{fig:Correlator_Soft}
  Examples for soft corrections to the non-factorizable part of the 
  three-point correlator~\eqref{eq:DefK}. The red, thick light-quark 
  line carries momentum of the order of $m_s\ll\omega\sim\Lbar$.}
 \end{center}
\end{figure}

We compute the $m_s$-expanded result for the leading non-factorizable part of the 
three-point correlators using the expansion by regions~\cite{Beneke:1997zp,Jantzen:2011nz}. 
As in the case of the two-point correlator, contributions involving soft propagators 
like the ones shown in Figure~\ref{fig:Correlator_Soft} first contribute at order $m_s^3$. 
Thus, we only have to consider the fully hard momentum region where all loop momenta 
admit the scaling $l\sim\omega_i\gg m_s$ and the loop integrands can be naively Taylor 
expanded in $m_s$. We have performed two independent calculations. The amplitudes are 
either generated using \texttt{QGRAF}~\cite{Nogueira:1991ex} with further processing in 
\texttt{Mathematica} or with a manual approach. The Dirac algebra is performed either 
with \texttt{TRACER}~\cite{Jamin:1991dp} or a private implementation.  We employ 
\texttt{FIRE}~\cite{Smirnov:2014hma} to generate IBP relations~\cite{Chetyrkin:1981qh} 
between the loop integrals and to reduce them to a set of Master integrals with the 
Laporta algorithm~\cite{Laporta:2001dd}. The required master integrals have been 
computed to all orders in $\epsilon$ in~\cite{Grozin:2008nu}. We have expanded them 
up to the required order in $\epsilon$ using \texttt{HypExp}~\cite{Huber:2007dx}. 
For completeness we state the results $r_{\tilde{Q}_i}^{(0)} = r_{\tilde{Q}_i}^{(0)}(x,L_\omega)$ 
for $m_s=0$ previously presented in~\cite{Kirk:2017juj}
\begin{eqnarray}
 r_{\tilde{Q}_1}^{(0)} & = & 8-\frac{a_2}{2}-\frac{8 \pi ^2}{3},\nonumber\\
 r_{\tilde{Q}_2}^{(0)} & = & 25+\frac{a_1}{2}-\frac{4 \pi ^2}{3}+6 L_\omega+\phi(x),\nonumber\\
 r_{\tilde{Q}_4}^{(0)} & = & 16-\frac{a_3}{4}-\frac{4 \pi ^2}{3}+3L_\omega+\frac{\phi(x)}{2},\nonumber\\
 r_{\tilde{Q}_5}^{(0)} & = & 29-\frac{a_3}{2}-\frac{8 \pi ^2}{3}+6 L_\omega+\phi(x),
\end{eqnarray}
with 
\begin{equation}
 \phi(x)=\begin{cases}
  x^2-8 x+6 \ln(x),\hspace{1cm}x\leq1,\\ 
  \frac{1}{x^2}-\frac{8}{x}-6 \ln(x),\hspace{1.2cm}x>1.
 \end{cases}
\end{equation}
For the linear terms $r_{\tilde{Q}_i}^{(1)} = r_{\tilde{Q}_i}^{(1)}(x,L_\omega)$ we obtain 
\begin{eqnarray}
 r_{\tilde{Q}_1}^{(1)} & = & -\frac{a_2}{2} - \frac{8\pi^2}{3} - 2\psi(x) + \begin{cases}
       \frac{2 (18-63 x+23 x^2)}{9 (1+x)}   + \left(2 - \frac{2 (3+x^3)}{3 x (1+x)}  \right)\ln(x), & x\leq1,\\
       \frac{2 (23-63 x+18 x^2)}{9 x (1+x)} - \left(2 - \frac{2 (1+3 x^3)}{3 x (1+x)}\right)\ln(x), & x>1,
       \end{cases}\nonumber\\
 r_{\tilde{Q}_2}^{(1)} & = & \frac{a_1}{2}-\frac{4 \pi ^2}{3}+6 L_\omega+\psi(x) + \begin{cases}
       \frac{243+162 x-41 x^2}{9 (1+x)}   + \left(5 + \frac{3+x^3}{3 x (1+x)}  \right)\ln(x), & x\leq1,\\
       \frac{243 x^2+162 x-41}{9 x (1+x)} - \left(5 + \frac{1+3 x^3}{3 x (1+x)}\right)\ln(x), & x>1,
       \end{cases}\nonumber\\
 r_{\tilde{Q}_4}^{(1)} & = & -\frac{a_3}{4}-\frac{4 \pi ^2}{3}+3L_\omega + \begin{cases}
       \frac{4 (36+9 x+x^2)}{9 (1+x)}     + \left(3 - \frac{2 x^2}{3 (1+x)}\right)\ln(x), & x\leq1,\\
       \frac{4 (1+9 x+36 x^2)}{9 x (1+x)} - \left(3 - \frac{2}{3 x (1+x)}  \right)\ln(x), & x>1,
       \end{cases}\nonumber\\
 r_{\tilde{Q}_5}^{(1)} & = & -\frac{a_3}{2}-\frac{8 \pi ^2}{3}+6 L_\omega + \begin{cases}
       \frac{29+11 x-2 x^2}{1+x}     + 6\ln(x), & x\leq1,\\
       \frac{29 x^2+11 x-2}{x (1+x)} - 6\ln(x), & x>1,
       \end{cases}
\end{eqnarray}
with 
\begin{equation}
 \psi(x) = \begin{cases}
            \frac{(1-x)^2}{x}\left[2\ln(1-x) - \ln(x)\right], \hspace{1cm}x\leq1,\\
            \frac{(1-x)^2}{x}\left[2\ln(x-1) - \ln(x)\right], \hspace{1cm}x>1.
           \end{cases}
\end{equation}
Last but not least, our results for the quadratic terms $r_{\tilde{Q}_i}^{(2)} = r_{\tilde{Q}_i}^{(2)}(x,L_\omega)$ are
\begin{eqnarray}
 r_{\tilde{Q}_1}^{(2)} & = & \frac{1}{1+x^2}\,\Bigg[
       \frac{(1-x)^2 a_2}{4} + \frac{2\pi^2(1-4x+x^2)}{3} + 2x\psi(x)\left(2+\frac{1+x}{1-x}\ln(x)\right)\nonumber\\ 
       & & \hspace{-0.7cm}+ \begin{cases}
       -\frac{2(6+6x-x^2+2x^3)}{3}  + 2(2-4x+x^2)\ln(x)  - 4(1-x^2)\text{Li}_2(1-1/x), & x\leq1,\\
       -\frac{2(2-x+6x^2+6x^3)}{3x} - 2(1-4x+2x^2)\ln(x) + 4(1-x^2)\text{Li}_2(1-x), & x>1,
       \end{cases}
       \Bigg],\nonumber\\ 
  r_{\tilde{Q}_2}^{(2)} & = & \frac{1}{1+x^2}\,\Bigg[
       \frac{-(1-x)^2 a_1}{4} - 3(1-x)^2L_\omega + \frac{\pi^2(1-4x+x^2)}{3} + \frac{x(1+x)}{1-x}\ln(x)\psi(x)\nonumber\\ 
       & & \hspace{-0.7cm}+ \begin{cases}
       -\frac{75-198x+89x^2-4x^3}{6}  - (3-6x+2x^2)\ln(x)  - 2(1-x^2)\text{Li}_2(1-1/x), & x\leq1,\\
       +\frac{4-89x+198x^2-75x^3}{6x} + (2-6x+3x^2)\ln(x) + 2(1-x^2)\text{Li}_2(1-x), & x>1,
       \end{cases}
       \Bigg],\nonumber\\ 
  r_{\tilde{Q}_4}^{(2)} & = & \frac{1}{1+x^2}\,\Bigg[
       \frac{(1-x)^2 a_3}{8} - \frac{3(1-x)^2}{2}L_\omega + \frac{x\psi(x)}{2}\left(1+\frac{3(1+x)}{1-x}\ln(x)\right)\nonumber\\ 
       & & + \begin{cases}
       -(1+8x-5x^2)\frac{\pi^2}{6} - \frac{24-48x+16x^2+x^3}{3}    - (1+x^2)\ln(x) \\
       - (1-x^2)\ln^2(x) - 5(1-x^2)\text{Li}_2(1-1/x), & x\leq1,\\
       +(5-8x-x^2)\frac{\pi^2}{6}  - \frac{1+16x-48x^2+24x^3}{3x}  + (1+x^2)\ln(x) \\
       + (1-x^2)\ln^2(x) + 5(1-x^2)\text{Li}_2(1-x), & x>1,
       \end{cases}
       \Bigg],\nonumber\\ 
  r_{\tilde{Q}_5}^{(2)} & = & \frac{1}{1+x^2}\,\Bigg[
       \frac{(1-x)^2 a_3}{4} - 3(1-x)^2 L_\omega + \frac{2\pi^2(1-4x+x^2)}{3}\nonumber\\ 
       & &  + 2x\psi(x)\left(1+\frac{1+x}{1-x}\ln(x)\right) - \frac{29-62x+29x^2}{2}\nonumber\\
       & &  + \begin{cases}
       - (1-x)^2\ln(x) - 4(1-x^2)\text{Li}_2(1-1/x), & x\leq1,\\
       + (1-x)^2\ln(x) + 4(1-x^2)\text{Li}_2(1-x), & x>1,
       \end{cases}
       \Bigg].\nonumber\\ 
\end{eqnarray}


\section{\boldmath Results and phenomenology\label{sec:results}}

We determine the Bag parameters in Section~\ref{sec:results_results}, give our predictions 
for the $B_s$ mixing observables in Section~\ref{sec:results_mixing} and use the results to 
determine the CKM elements $|V_{td}|$ and $|V_{ts}|$ in Section~\ref{sec:ckm} and the top-quark 
$\overline{\text{MS}}$ mass in Section~\ref{sec:mt}. We then present an alternative prediction 
of the branching ratios $\mathcal{B}(B_{q}\to\mu^+\mu^-)$ from the ratios 
$\mathcal{B}(B_{q}\to\mu^+\mu^-)/\Delta M_q$ in Section~\ref{sec:bmumu}.
Our analysis strategy closely follows the one we used in~\cite{Kirk:2017juj} in the 
limit $m_s=0$ and we only comment on where they differ due to the non-zero strange mass 
while referring to~\cite{Kirk:2017juj} for more details.


\subsection{Bag parameters\label{sec:results_results}}

We determine the HQET Bag parameters at the scale $\mu_\rho=1.5$ GeV using the weight 
function approach \eqref{eq:B_NoInt}. The strange-quark mass scheme in \eqref{eq:B_NoInt} 
is undetermined since any scheme change would only affect the expressions at higher orders 
which are not taken into account. We use the value in the $\overline{\text{MS}}$ scheme at 
the scale $\mu_\rho$ which is determined from the central value of the average 
$\overline{m}_s(2\,\text{GeV}) = (95_{-3}^{+9})\,$MeV~\cite{Tanabashi:2018oca}. 
To account for the uncertainties related to the scheme choice and the truncation of the 
expansion in $m_s$ we increase the parametric uncertainty and use 
$\overline{m}_s(2\,\text{GeV}) = (95\pm30)\,$MeV. 
To the perturbative part we add the condensate contributions~\cite{Mannel:2007am,Mannel:2011zza}. 
The lattice simulation \cite{McNeile:2012xh} shows that light and strange quark condensates 
agree within uncertainties and their result for the strange-quark condensate has since been 
confirmed with a different method \cite{Davies:2018hmw}. With the factorization hypothesis 
$\langle\bar{q}Gq\rangle = m_0^2\langle\bar{q}q\rangle$ the same holds for the quark-gluon 
condensate. We therefore assume the condensate corrections to be the same in the $B^0$ and 
$B_s^0$ systems. We obtain 
\begin{align}
 B_{\tilde{Q}_1}^s (1.5\text{ GeV}) & = (0.910-0.016_{m_s}+0.003_{m_s^2})\,_{-0.036}^{+0.025} \nonumber\\
                                    & = 0.897\,_{-0.002}^{+0.002}(\Lbar)\,_{-0.020}^{+0.020}(\text{intr.})\,_{-0.005}^{+0.005}(\text{cond.})
                                        \,_{-0.029}^{+0.014}(\mu_\rho)\,_{-0.003}^{+0.003}(m_s),\vspace*{0.1cm}\nonumber\\
 B_{\tilde{Q}_2}^s (1.5\text{ GeV}) & = (0.939-0.006_{m_s}+0.002_{m_s^2})\,_{-0.031}^{+0.027} \nonumber\\
                                    & = 0.936\,_{-0.016}^{+0.014}(\Lbar)\,_{-0.020}^{+0.020}(\text{intr.})\,_{-0.004}^{+0.004}(\text{cond.})
                                        \,_{-0.016}^{+0.011}(\mu_\rho)\,_{-0.004}^{+0.004}(m_s),\vspace*{0.1cm}\nonumber\\
 B_{\tilde{Q}_4}^s (1.5\text{ GeV}) & = (1.003-0.004_{m_s}+0.001_{m_s^2})\,_{-0.023}^{+0.023} \nonumber\\
                                    & = 1.000\,_{-0.004}^{+0.005}(\Lbar)\,_{-0.020}^{+0.020}(\text{intr.})\,_{-0.010}^{+0.010}(\text{cond.})
                                        \,_{-0.002}^{+0.000}(\mu_\rho)\,_{-0.002}^{+0.003}(m_s),\vspace*{0.1cm}\nonumber\\
 B_{\tilde{Q}_5}^s (1.5\text{ GeV}) & = (0.988-0.008_{m_s}+0.000_{m_s^2})\,_{-0.027}^{+0.028} \nonumber\\
                                    & = 0.980\,_{-0.012}^{+0.015}(\Lbar)\,_{-0.020}^{+0.020}(\text{intr.})\,_{-0.010}^{+0.010}(\text{cond.})
                                        \,_{-0.007}^{+0.000}(\mu_\rho)\,_{-0.006}^{+0.007}(m_s),
 \label{eq:DelB2_HQET_results_Bs}
\end{align}
where we have indicated the orders in $m_s$ with subscripts and find good convergence of 
the expansion. The differences in the leading terms with respect to the results for $B_d$ mixing 
obtained in \cite{Kirk:2017juj} arise because the logarithms $L_{\Lbar}$ are replaced by 
$L_{\Lbar+m_s}$ which we do not expand in $m_s/\Lbar$. 

\begin{figure}[t]
 \begin{center}
    \includegraphics[width=0.7\textwidth]{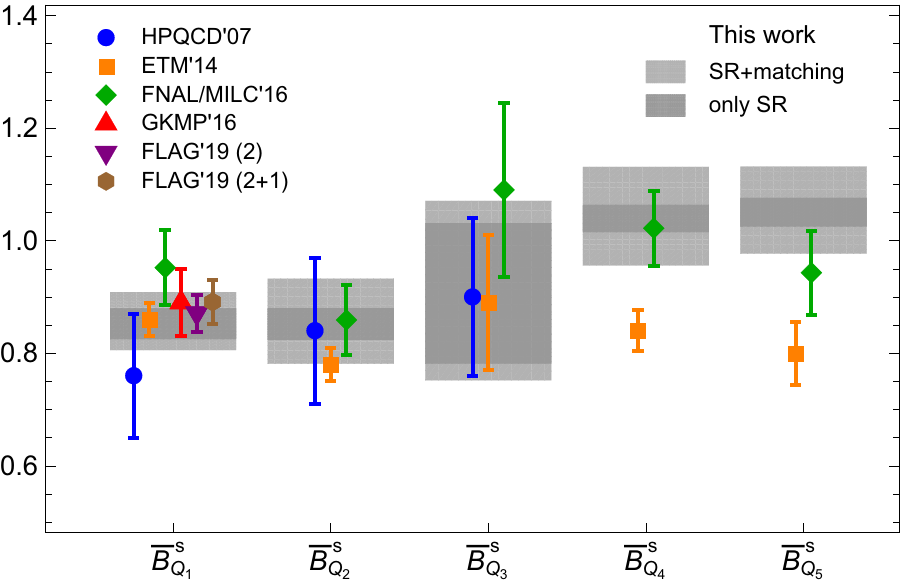}
  \caption{
  Comparison of Bag parameters relevant for $B_s$ mixing. The dark gray regions indicate the ranges 
  spanned only by the sum rule error whereas the light gray regions correspond to the total uncertainties. 
  The sum rule value GKMP'16 corresponds to the result~\cite{Grozin:2016uqy} for the $B_d$ system with an 
  uncertainty of $\pm0.02$ for the $m_s$ effects added in quadrature as suggested by the authors in~\cite{Grozin:2017uto}. 
  }
  \label{fig:Bag_Comparison}
 \end{center}
\end{figure}

The results \eqref{eq:DelB2_HQET_results_Bs} are then evolved to the matching scale 
$\mu_m=\overline{m}_b(\overline{m}_b)$ where they are converted to QCD Bag parameters 
$B_Q^s$ using \eqref{eq:B_QCD_HQET_matching}. We do not consider the effects of a non-zero 
strange-quark mass in the QCD-HQET matching. The matching corrections are of the order 
$\alpha_s(\overline{m}_b(\overline{m}_b))/\pi\times\overline{m}_s(\overline{m}_b)/\overline{m}_b(\overline{m}_b)\sim0.001$ 
and therefore subleading compared to the linear terms 
$\alpha_s(\mu_\rho)/\pi\times\overline{m}_s(\mu_\rho)/(\Lbar+\overline{m}_s(\mu_\rho))\sim0.019$ 
and even the quadratic terms 
$\alpha_s(\mu_\rho)/\pi\times[\overline{m}_s(\mu_\rho)/(\Lbar+\overline{m}_s(\mu_\rho))]^2\sim0.003$ 
in the sum rule. We do not include this uncertainty as a separate contribution in our error 
analysis since it is covered by the conservative variation of the input value for $m_s$. 
Lastly, we convert the QCD Bag parameters to the usual convention which we denoted as 
$\overline{B}_Q^s$ in \eqref{eq:Bags_QCD}. We find 
\begin{align}
 \overline{B}_{Q_1}^s(\overline{m}_b(\overline{m}_b)) & = 0.858_{-0.052}^{+0.051} 
      = (0.870 - 0.015_{m_s} + 0.002_{m_s^2})_{-0.033}^{+0.022}(\text{SR})_{-0.040}^{+0.046}(\text{M}),\nonumber\\
 \overline{B}_{Q_2}^s(\overline{m}_b(\overline{m}_b)) & = 0.854_{-0.072}^{+0.079}
      = (0.857 - 0.005_{m_s} + 0.002_{m_s^2})_{-0.030}^{+0.026}(\text{SR})_{-0.066}^{+0.074}(\text{M}),\nonumber\\
 \overline{B}_{Q_3}^s(\overline{m}_b(\overline{m}_b)) & = 0.907_{-0.155}^{+0.164}
      = (0.880 + 0.027_{m_s} + 0.000_{m_s^2})_{-0.125}^{+0.124}(\text{SR})_{-0.091}^{+0.107}(\text{M}),\nonumber\\
 \overline{B}_{Q_4}^s(\overline{m}_b(\overline{m}_b)) & = 1.039_{-0.083}^{+0.092}
      = (1.043 - 0.004_{m_s} + 0.001_{m_s^2})_{-0.024}^{+0.024}(\text{SR})_{-0.080}^{+0.088}(\text{M}),\nonumber\\
 \overline{B}_{Q_5}^s(\overline{m}_b(\overline{m}_b)) & = 1.050_{-0.074}^{+0.081}
      = (1.058 - 0.007_{m_s} + 0.000_{m_s^2})_{-0.025}^{+0.025}(\text{SR})_{-0.069}^{+0.077}(\text{M}), 
 \label{eq:DelB2_QCD_results_Bs}
\end{align}
where we have included the uncertainty from variation of $\overline{m}_s$ in the sum rule (SR) 
error and M denotes the uncertainty from the QCD-HQET matching. 
We compare our results to other determinations from lattice simulations~\cite{Dalgic:2006gp,Carrasco:2013zta,Bazavov:2016nty} 
and sum rules~\cite{Grozin:2016uqy} and the FLAG averages~\cite{Aoki:2019cca} in Figure~\ref{fig:Bag_Comparison} 
and find very good agreement overall with similar uncertainties. We observe that the FNAL/MILC'16 value for 
$\overline{B}_{Q_1}$ is larger than all the other results -- with respect to our value the difference corresponds 
to 1.1 sigma. We note that FNAL/MILC'16 determined the combination $f_{B_s}^2\overline{B}_{Q_1}$ and 
extracted the Bag parameter using the 2016 PDG average for the decay constant. They are currently working on 
a direct determination and, since their recent result \cite{Bazavov:2017lyh} for $f_{B_s}$ is larger than the 
PDG value used in \cite{Bazavov:2016nty}, we expect the Bag parameter to go down. On the other hand our Bag parameters for 
$Q_{4,5}$ are in good agreement with FNAL/MILC'16, while there is a tension of more than two sigmas with respect 
to the results of ETM'14. Similar tensions have been observed in the Kaon system~\cite{Boyle:2017ssm} 
where it was conjectured that a difference in intermediate renormalization schemes might be responsible. 
We also consider the ratios $\overline{B}_{Q_1}^{s/d} \equiv \overline{B}_{Q_1}^s/\overline{B}_{Q_1}^d$ of the 
Bag parameters in the $B_s^0$ and $B_d^0$ system where a large part of the uncertainties cancel 
\begin{align}
 \overline{B}_{Q_1}^{s/d}(\overline{m}_b(\overline{m}_b)) & = 0.987_{-0.009}^{+0.007} = (1.001-0.017_{m_s}+0.003_{m_s^2})_{-0.008}^{+0.007}(\text{SR})_{-0.002}^{+0.002}(\text{M}),\nonumber\\
 \overline{B}_{Q_2}^{s/d}(\overline{m}_b(\overline{m}_b)) & = 1.013_{-0.008}^{+0.010} = (1.017-0.006_{m_s}+0.002_{m_s^2})_{-0.008}^{+0.009}(\text{SR})_{-0.002}^{+0.002}(\text{M}),\nonumber\\
 \overline{B}_{Q_3}^{s/d}(\overline{m}_b(\overline{m}_b)) & = 1.108_{-0.051}^{+0.068} = (1.076+0.033_{m_s}-0.001_{m_s^2})_{-0.051}^{+0.068}(\text{SR})_{-0.007}^{+0.007}(\text{M}),\nonumber\\
 \overline{B}_{Q_4}^{s/d}(\overline{m}_b(\overline{m}_b)) & = 0.991_{-0.008}^{+0.007} = (0.994-0.004_{m_s}+0.001_{m_s^2})_{-0.008}^{+0.006}(\text{SR})_{-0.002}^{+0.002}(\text{M}),\nonumber\\
 \overline{B}_{Q_5}^{s/d}(\overline{m}_b(\overline{m}_b)) & = 0.979_{-0.014}^{+0.010} = (0.985-0.007_{m_s}+0.000_{m_s^2})_{-0.013}^{+0.010}(\text{SR})_{-0.002}^{+0.002}(\text{M}).
 \label{eq:DelB2_QCD_results_ratio}
\end{align}
The leading terms in the $m_s$-expansion differ from unity because we do 
not expand the logarithms $L_{\Lbar+m_s}$ in $m_s/\Lbar$. 
Compared to the absolute Bag parameters we reduce the intrinsic sum rule error 
to 0.005, the condensate error to 0.002 and the uncertainty due to power corrections 
to 0.002 since the respective uncertainties cancel to a large extend in the ratios. 
However, we enhance the intrinsic sum rule and condensate error estimates for the 
operator $Q_3$ by a factor of five since the sum rule uncertainties for this operator 
are enhanced by large ratios of color factors $A_{Q_{1,2}}/A_{Q_3}$ as discussed 
in~\cite{Kirk:2017juj}. A detailed overview of the uncertainties is given in Appendix~\ref{sec:Uncertainties}. 
The ratios~\eqref{eq:DelB2_QCD_results_ratio} are in excellent agreement with the 
parametric estimates $1\pm0.02$ from \cite{Grozin:2017uto,Kirk:2017juj} with the exception 
of $Q_3$ where this uncertainty should have been enhanced like the other sum rule 
uncertainties listed above to account for the large color factors in the QCD-HQET matching 
relation~\eqref{eq:B_QCD_HQET_matching} for the Bag parameter. 

\begin{figure}[t]
 \begin{center}
    $\vtop{\vskip0pt\hbox{\includegraphics[width=0.49\textwidth]{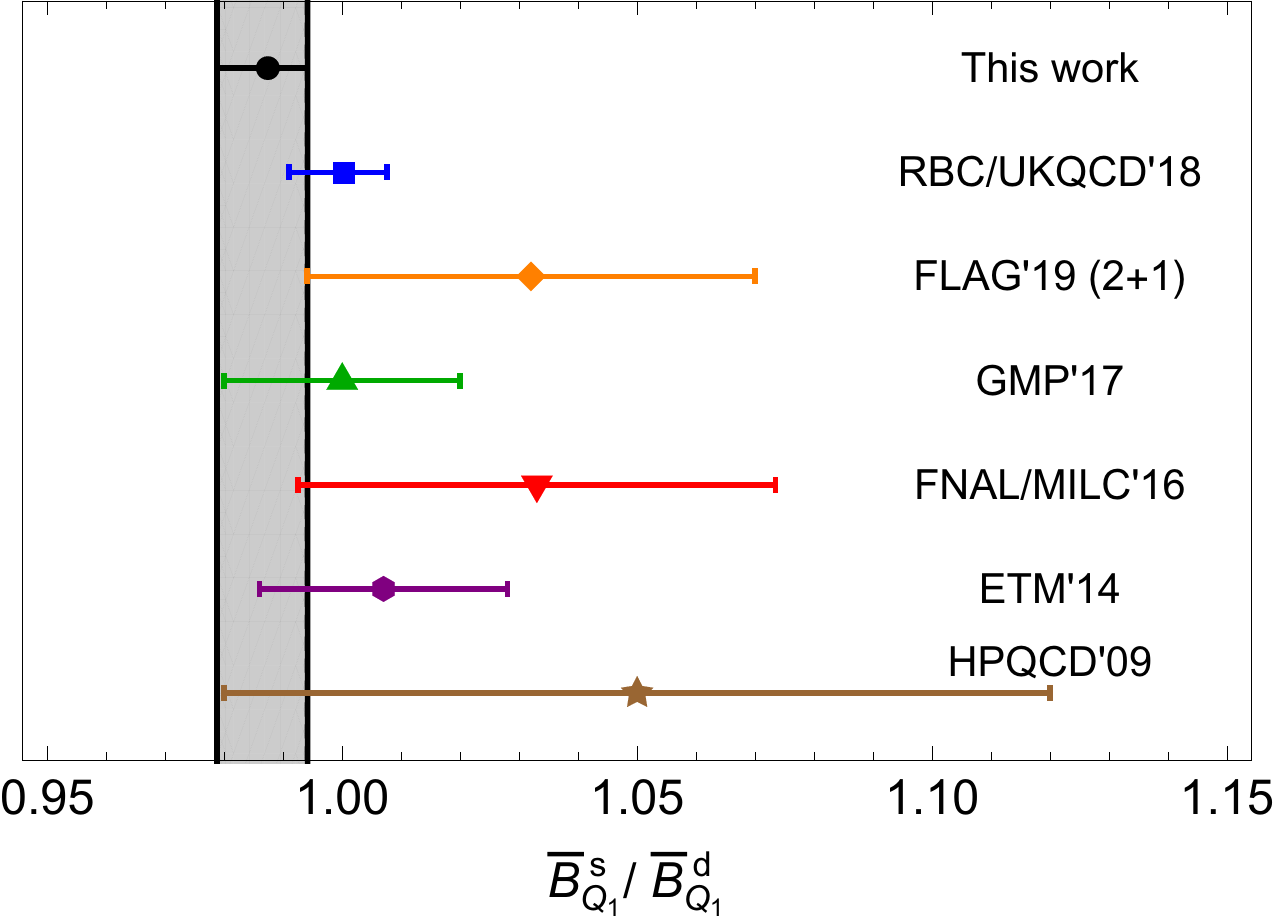}}} \hfill 
    \vtop{\vskip0pt\hbox{\includegraphics[width=0.49\textwidth]{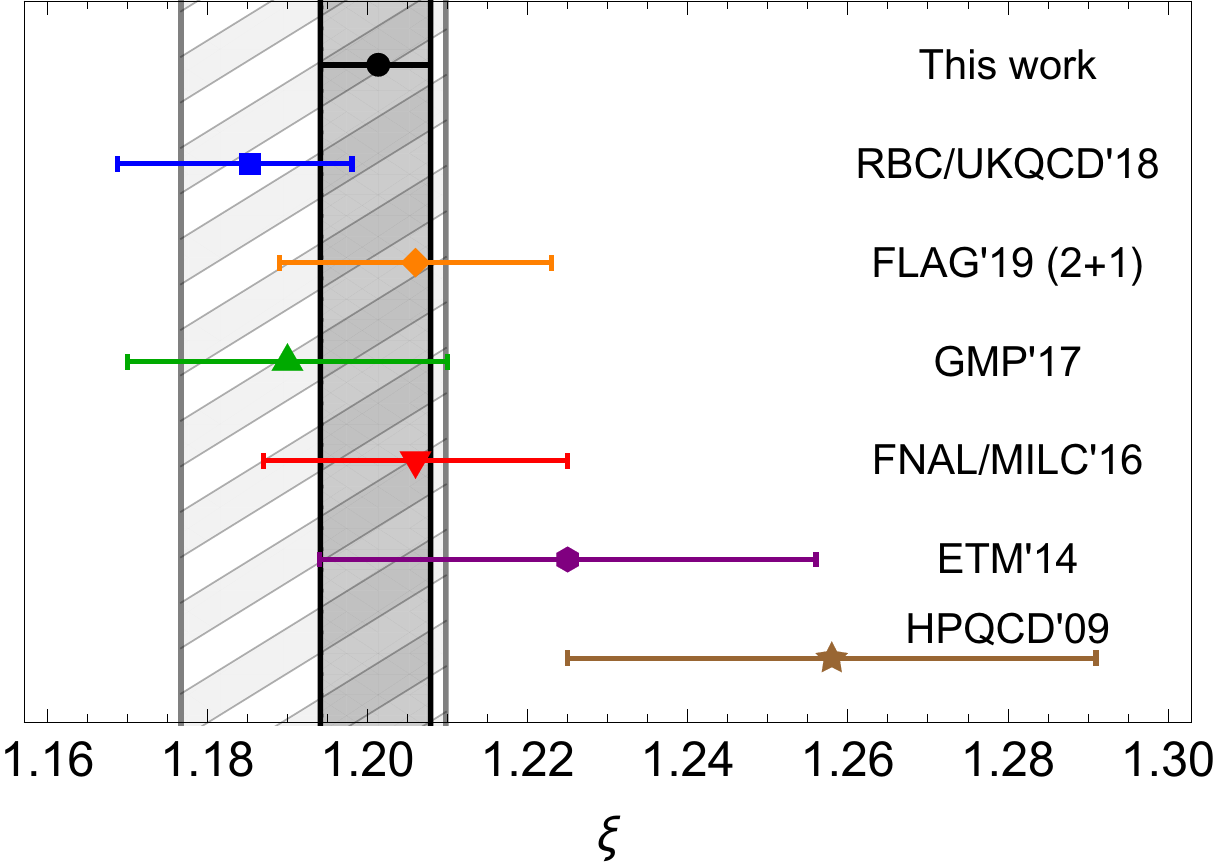}}}$
  \caption{
  Comparison of the ratios $\overline{B}_{Q_1}^s/\overline{B}_{Q_1}^d$ and $\xi$ defined in 
  \eqref{eq:XiResult} with results from the literature~\cite{Gamiz:2009ku,Carrasco:2013zta,Bazavov:2016nty,Grozin:2017uto,Boyle:2018knm,Aoki:2019cca}. 
  On the right side we also show our result obtained using the FLAG $N_f=2+1$ average for 
  the ratio of the decay constants as a hatched band. The GMP'17~\cite{Grozin:2017uto} value 
  for $\xi$ corresponds to Eq. (5.9) of that paper where the world average for $f_{B_s}/f_B$ 
  is used. 
  }
  \label{fig:Bag_Ratio_Comparison}
 \end{center}
\end{figure}

Taking the FLAG~\cite{Aoki:2019cca}\footnote{The average is dominated by the 
HPQCD'17~\cite{Hughes:2017spc} and FNAL/MILC'17~\cite{Bazavov:2017lyh} results.} 
value with $N_f=2+1+1$ for the ratio $f_{B_s}/f_B$ of the decay constants of $B_s^0$ 
and $B_d^0$ we obtain the most precise result to date for the ratio 
\begin{equation}
 \xi \equiv \frac{f_{B_s}}{f_B}\,\sqrt{\overline{B}_{Q_1}^{s/d}} = 1.2014_{-0.0072}^{+0.0065} 
          = 1.2014\pm0.0050\left(\frac{f_{B_s}}{f_B}\right)\,_{-0.0053}^{+0.0043}\left(\overline{B}_{Q_1}^{s/d}\right)\,,
\label{eq:XiResult}
\end{equation}
where the ratio of decay constants and Bag parameters contributes equally to the 
error budget. A comparison with previous results is shown in Figure~\ref{fig:Bag_Ratio_Comparison}. 
There we also show how the result changes when the FLAG $N_f=2+1$ average is used for the 
ratio of the decay constants. Unfortunately FNAL/MILC and ETM do not provide 
values for $\overline{B}_{Q_i}^{s/d}$ for $i=2,3,4,5$ so we cannot easily compare our
results for these ratios.


\subsection{\boldmath$B_s$ mixing observables\label{sec:results_mixing}}

\begin{figure}[t]
 \begin{center}
    $\vtop{\vskip0pt\hbox{\includegraphics[width=0.50\textwidth]{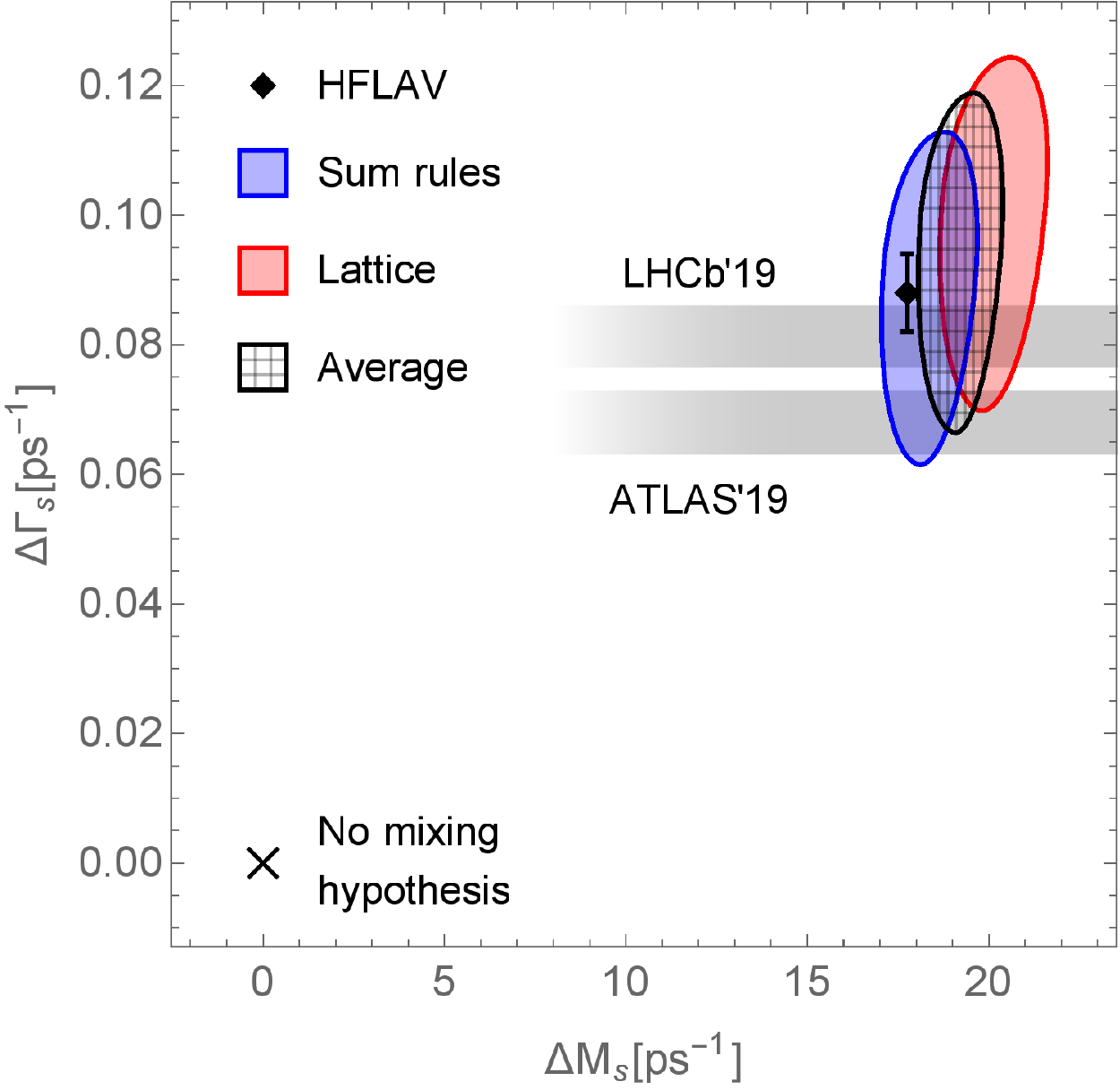}}}\hfill
     \vtop{\vskip0pt\hbox{\includegraphics[width=0.48\textwidth]{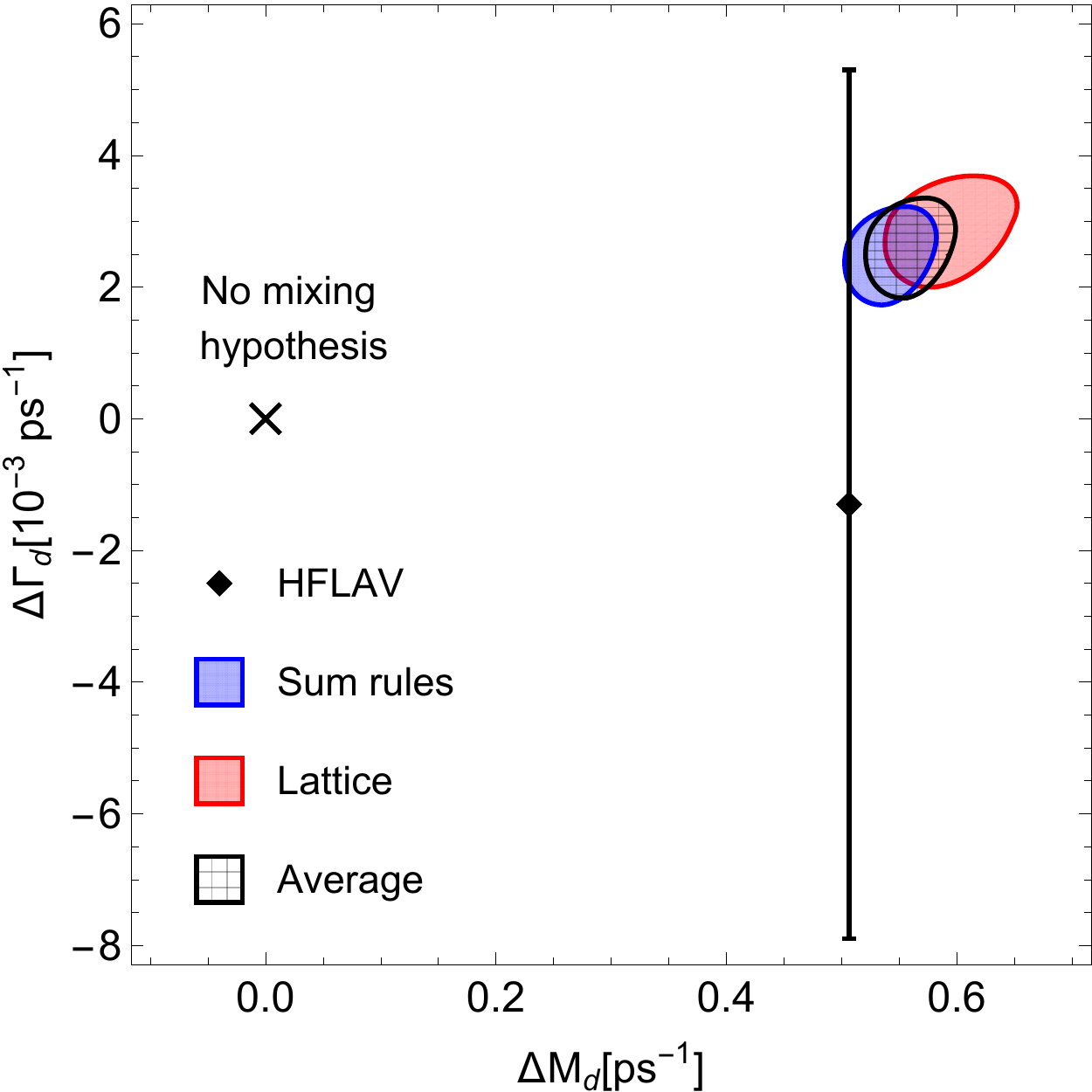}}}$
  \caption{
Our predictions (blue) for the mass and decay rate difference in the $B_s$ (left) and $B_d$ 
(right) systems are compared to the current experimental averages and the predictions (red) based 
on the latest lattice averages from FLAG~\cite{Aoki:2019cca} for $f_{B_q}^2\overline{B}_{Q_1}^q$ 
and the FNAL/MILC'16~\cite{Bazavov:2016nty} results for $f_{B_q}^2\overline{B}_{Q_i}^q$ with 
$i\neq1$ and $\langle R_0\rangle$. The weighted average over the sum rule and lattice results 
is shown in black. We indicate the updated Run 1 and Run 2 combinations for $\Delta \Gamma_s$ 
presented by LHCb~\cite{LHCb:2019Moriond} and ATLAS~\cite{ATLAS:2019Moriond} at Moriond EW 2019 
by shaded gray regions.}
  \label{fig:MixingSMvsExp}
 \end{center}
\end{figure}

In this section we present the results of our $B$ mixing analysis. We consider the mass differences 
$\Delta M_s$ and $\Delta M_d$, the decay rate differences $\Delta\Gamma_s$ and $\Delta\Gamma_d$, 
and the ratio $\Delta M_s/\Delta M_d$, of which the latter benefits from a reduced uncertainty 
due to the cancellation of CKM factors and hadronic effects. For the bottom-quark mass we studied
the $\overline{\text{MS}}$, PS~\cite{Beneke:1998rk}, 1S~\cite{Hoang:1998ng} and the 
kinetic~\cite{Bigi:1996si} mass schemes and found good agreement (see \cite{Kirk:2017juj} for a 
more detailed discussion) - below we just quote the result in the PS scheme. 
We choose as our CKM parameter inputs 
the results of CKMfitter2018\cite{Charles:2004jd} and collect these along with our other 
numerical inputs in Appendix \ref{sec:Uncertainties}. For the non-perturbative input we use our SR
determination of the Bag parameters (Eq.(\ref{eq:DelB2_QCD_results_Bs}) and Eq.
(\ref{eq:DelB2_QCD_results_ratio})) together with the lattice decay constants ($N_f = 2+1+1$) from
\cite{Aoki:2019cca} (dominated by HPQCD'17~\cite{Hughes:2017spc} and FNAL/MILC'17~\cite{Bazavov:2017lyh}).
\\
Comparing our findings for $\Delta M_s$ 
we see an excellent agreement with the experimental measurement \cite{Amhis:2016xyh}: 
\begin{eqnarray}
 \Delta M_s^\text{exp}  & = &(17.757 \pm 0.021)\,\text{ps}^{-1},
 \nonumber
 \\
 \Delta M_s^\text{SR}   & = & (18.5_{-1.5}^{+1.2})\,\text{ps}^{-1} 
 \nonumber
 \\
                        & = & (18.5\pm1.1\,(\text{had.})\pm 0.1\,(\text{scale})_{-1.0}^{+0.3}\,\,(\text{param.}))\,\text{ps}^{-1} \, .
\end{eqnarray}
We note that the update to our CKM input gives rise to an increase in $\Delta M_s^{\text{SR}}$ 
from the value presented in \cite{Kirk:2017juj}, despite the inclusion of $m_s$-corrections 
which reduce the size of our hadronic input. Using instead the non-perturbative input purely from lattice
determinations (FLAG 2019 \cite{Aoki:2019cca}, which is almost identical to the result in \cite{Bazavov:2016nty}), 
we get a considerably higher SM prediction for $\Delta M_s$: 
$\Delta M_s^\text{Lat.}  = (20.3_{-1.7}^{+1.3})\,\text{ps}^{-1} 
 = (20.3\pm1.3\,(\text{had.})\pm 0.1\,(\text{scale})_{-1.1}^{+0.3}\,\,(\text{param.}))\,\text{ps}^{-1}$,
 being about 1.5 standard deviations above the experiment.
 Due to updated CKM inputs this number is slightly larger than the one quoted in Eq.(\ref{DeltaMsSM1}).
Averaging the SR and the lattice results, we get a further reduction of the uncertainties:
$\Delta M_s^\text{Av.}   = (19.4_{-1.4}^{+1.0})\,\text{ps}^{-1} 
 = (19.4\pm0.9\,(\text{had.})\pm 0.1\,(\text{scale})_{-1.0}^{+0.3}\,\,(\text{param.}))\,\text{ps}^{-1}$.
\\
We also find perfect agreement between our result for $\Delta\Gamma_s$ and experiment \cite{Amhis:2016xyh}:
\begin{eqnarray}
 \Delta \Gamma_s^\text{exp}  & = &(0.088 \pm 0.006)\,\text{ps}^{-1},
 \nonumber
 \\
  \Delta \Gamma_s^\text{SR}   & = & (0.091_{-0.030}^{+0.022})\,\text{ps}^{-1} 
  \nonumber
  \\
& = & (0.091\pm 0.020\,(\text{had.})_{-0.021}^{+0.008}\,\,(\text{scale})_{-0.005}^{+0.002}\,\,(\text{param.}))\,\text{ps}^{-1}.
\end{eqnarray}
Recent measurements~\cite{LHCb:2019Moriond,ATLAS:2019Moriond} that are not yet contained in 
the average \cite{Amhis:2016xyh} yield significantly smaller values for $\Delta\Gamma_s$ which 
are however still in the one-sigma range of our prediction. 
The theoretical prediction for the decay rate difference includes NLO QCD 
\cite{Beneke:1998sy,Beneke:2003az,Ciuchini:2003ww,Lenz:2006hd} 
and $1/m_b$  \cite{Beneke:1996gn,Dighe:2001gc} corrections. 
The latter depend on matrix elements of dimension-seven operators which are currently only known 
in the vacuum saturation approximation, which results in uncertainties of approximately 25-30\text{\%}. 
The sizable scale uncertainty can be reduced with a NNLO computation of the HQE matching 
coefficients - first steps towards this have recently been performed in~\cite{Asatrian:2017qaz}. 
Using instead the non-perturbative input from lattice \cite{Aoki:2019cca}, we again get higher 
values $\Delta \Gamma_s^\text{Lat.} = (0.102_{-0.032}^{+0.023})\,\text{ps}^{-1}  = 
(0.102\pm 0.020\,(\text{had.})_{-0.024}^{+0.010}\,\,(\text{scale})_{-0.006}^{+0.002}\,\,(
\text{param.}))\text{ps}^{-1}$. Due to the larger uncertainties this prediction overlaps at 1 
sigma with experiment. Combining the the sum rule result with the lattice result we get
$\Delta \Gamma_s^\text{Av.}   = (0.097_{-0.031}^{+0.022})\,\text{ps}^{-1} 
 = (0.097\pm 0.020\,(\text{had.})_{-0.023}^{+0.009}\,\,(\text{scale})_{-0.005}^{+0.002}\,\,
 (\text{param.}))\,\text{ps}^{-1}$. Here the accuracy of the average does not improve, because the uncertainty
 is dominated by the unknown matrix elements of dimension seven operators and scale variation. 
\\
Due to new CKM inputs (compared to the $B_d$ analysis in \cite{Kirk:2017juj}), we are also updating our results for $B_d$ mixing observables\footnote{The corresponding lattice result reads
$\Delta M_d^\text{Lat.}  = (0.596_{-0.063}^{+0.054})\,\text{ps}^{-1}  $
(about 1.4 sigma above experiment)
and the average over SR and lattice is
$\Delta M_d^\text{Av.}  = (0.565_{-0.046}^{+0.034})\,\text{ps}^{-1} $
.}:
\begin{eqnarray}
 \Delta M_d^\text{exp}  & = &(0.5064 \pm 0.0019)\,\text{ps}^{-1},
 \nonumber
 \\
 \Delta M_d^\text{SR}   & = & (0.547_{-0.046}^{+0.035})\,\text{ps}^{-1} 
 \nonumber\\
 & = & (0.547_{-0.032}^{+0.033}\,(\text{had.})_{-0.002}^{+0.004}\,(\text{scale})_{-0.032}^{+0.011}\,\,(
 \text{param.}))\,\text{ps}^{-1},
 \end{eqnarray}
and\footnote{The corresponding lattice result reads
$ \Delta \Gamma_d^\text{Lat.} = (3.0_{-1.0}^{+0.7})\cdot10^{-3}\,\text{ps}^{-1} $
and the average over SR and lattice is
$\Delta \Gamma_d^\text{Av.}  = (2.7_{-0.9}^{+0.6})\cdot10^{-3}\,\text{ps}^{-1} 
$.}:
\begin{eqnarray}
 \Delta \Gamma_d^\text{exp}  & = (& -1.3 \pm 6.6)\cdot10^{-3}\,\text{ps}^{-1},
 \nonumber
 \\
 \Delta \Gamma_d^\text{SR}   & = & (2.6_{-0.9}^{+0.6})\cdot10^{-3}\,\text{ps}^{-1} 
 \nonumber
 \\
& = & (2.6\pm 0.6\,(\text{had.})_{-0.6}^{+0.2}\,\,(\text{scale})_{-0.2}^{+0.1}\,\,(\text{param.}))\cdot10^{-3}\,\text{ps}^{-1} ,
\end{eqnarray}
where at present only an experimental upper bound on $\Delta\Gamma_d^{\text{exp}}$ is available.
The SM value of the mass difference agrees with experiment at the 1 sigma level.
Fig. \ref{fig:MixingSMvsExp} (left panel) shows the comparison of the measurements of $\Delta \Gamma_s$
and  $\Delta M_s$ with the corresponding theory predictions: in blue the 1 sigma region of our 
sum rule values, in the red the purely lattice results and in black the average of both. 
The right panel shows the same comparison for the $B_d$ system. All in all the sum rule values 
agree well with experiment, while the pure lattice results show a 1.5 sigma deviation for 
the mass differences - leading to very strong bounds on BSM models that try to explain the flavour 
anomalies. 

\noindent
Finally, for the ratio of the mass differences we also find our results to be consistent (within about 1.3
standard deviations) with the measured value:
\begin{eqnarray}
 \left(\frac{\Delta M_d}{\Delta M_s}\right)_\text{exp}  & = & 0.0285\pm0.0001, 
 \nonumber
 \\
 \left(\frac{\Delta M_d}{\Delta M_s}\right)_\text{SR}   & = & 0.0297_{-0.0009}^{+0.0006} = 0.0297_{-0.0003}^{+0.0004}\,(\text{had.})_{-0.0008}^{+0.0005}\,(\text{exp.}).
\end{eqnarray}
Due to our new value for $\xi$ we get a theoretical precision of about $3\%$ for the ratio of
mass differences in the $B_d$ and $B_s$ systems, which poses severe constraints on BSM models,
that modify neutral $B$ meson mixing.
The uncertainty is now dominated by the CKM factors. Using lattice inputs one gets a slightly less precise
value
$(\Delta M_d / \Delta M_s)_\text{Lat.} = 0.0295_{-0.0012}^{+0.0010}  =
0.0295_{-0.0008}^{+0.0008}\,(\text{had.})_{-0.0008}^{+0.0005}\,(\text{exp.})$,
which can be combined with the sum rule result to obtain
$  \left(\Delta M_d / \Delta M_s\right)_\text{Av.}   = 0.0297_{-0.0009}^{+0.0006}  =
0.0297_{-0.0003}^{+0.0003}\,(\text{had.})_{-0.0008}^{+0.0005}\,(\text{exp.}). $


\subsection{\boldmath Determination of the CKM elements $|V_{td}|$ and $|V_{ts}|$\label{sec:ckm}}
We also can use the measured values of the mass differences, together with our bag parameter, 
the lattice results for the decay constant ($N_f = 2+1+1$ from
\cite{Aoki:2019cca,Hughes:2017spc,Bazavov:2017lyh}) and the value of the CKM element $V_{tb}$ 
(from \cite{Charles:2004jd}) to determine
$|V_{td}|$ and $|V_{ts}|$
\begin{eqnarray}
 \left|V_{ts}\right|_\text{SR}  & = & (40.74_{-1.21}^{+1.30})\cdot10^{-3} 
 \nonumber\\
 & = & 
 (40.74_{-1.20}^{+1.29}\,(\text{had.})\,_{-0.14}^{+0.09}\,(\mu)\,\pm0.05\,(\text{param.}))\cdot10^{-3}\,,
 \nonumber\\
 \left|V_{td}\right|_\text{SR}  & = & ( 8.36_{-0.24}^{+0.26})\cdot10^{-3} 
 \nonumber
 \\
 &= &( 8.36_{-0.24}^{+0.26}\,(\text{had.})\,_{-0.03}^{+0.02}\,(\mu)\,\pm0.02\,(\text{param.}))\cdot10^{-3}\,.
\end{eqnarray}
These direct determinations overlap with the determinations based on CKM unitarity \cite{Charles:2004jd}
(see \cite{Bona:2006ah} for similar results) but they are a little less precise:
\begin{eqnarray}
 \left|V_{ts}\right|_\text{CKMfitter}  & = & (41.69_{-1.08}^{+0.28})\cdot10^{-3} 
 \nonumber\\
 \left|V_{td}\right|_\text{CKMfitter}  & = & ( 8.710_{-0.246}^{+0.086})\cdot10^{-3} \,.
\end{eqnarray}
We note that the results of the full CKM fit  include data on $B$ mixing and are therefore not 
completely independent. Thus, it is also interesting to compare to the results of the fit 
where only tree-level processes are considered. A discrepancy here would be a hint towards new 
physics in loop processes. The CKMfitter results are 
\begin{eqnarray}
 \left|V_{ts}\right|_\text{CKMfitter, tree}  & = & (41.63_{-1.45}^{+0.39})\cdot10^{-3} 
 \nonumber\\
 \left|V_{td}\right|_\text{CKMfitter, tree}  & = & ( 9.08_{-0.45}^{+0.23})\cdot10^{-3} \,.
\end{eqnarray}
While there is good agreement for $|V_{ts}|$ the value of $|V_{td}|$ differs from our result 
by about 1.4 sigma. The value of the ratio $|V_{td}/V_{ts}|$ can be determined more precisely 
based on the exact relation
\begin{equation}
     \frac{\Delta M_d}{\Delta M_s} = \left| \frac{V_{td}}{V_{ts}} \right|^2 \frac{1}{\xi^2}
     \frac{M_{B_d}}{M_{B_s}} \, .
\end{equation}
Using our value of $\xi$ from Eq.~\eqref{eq:XiResult} we can present here the most precise determination of
$|V_{td}/V_{ts}|$:
\begin{equation}
    \left|V_{td}/V_{ts}\right|_\text{SR} = 0.2045_{-0.0013}^{+0.0012} = 0.2045_{-0.0012}^{+0.0011}\,(\text{had.})\,\pm0.0004\,(\text{exp.})\,,\\
\label{eq:VtdVts_result}
\end{equation}
which is compatible with the values obtained by the FNAL/MILC \cite{Bazavov:2016nty} and 
RBC-UKQCD \cite{Boyle:2018knm} collaborations 
\begin{equation}
\begin{array}{ll}
 \left|V_{td}/V_{ts}\right| = 0.2052\pm0.0033  \hspace{1cm} &[\text{FNAL/MILC'16}] \, ,
 \\
 \left|V_{td}/V_{ts}\right| = 0.2018_{-0.0027}^{+0.0020} \hspace{1cm} &[\text{RBC-UKQCD'18}] \, .
\end{array}
\end{equation}
These values are all somewhat smaller than the expectation from CKM unitarity taken from 
CKMfitter \cite{Charles:2004jd} and UTfit \cite{Bona:2006ah} 
\begin{equation}
\begin{array}{ll}
 \left|V_{td}/V_{ts}\right| = 0.2088^{+0.0016}_{-0.0030} \hspace{1cm} &[\text{CKMfitter}] \, ,
 \\
 \left|V_{td}/V_{ts}\right| = 0.211\pm0.003      \hspace{1cm}     &[\text{UTfit}] \, .
\end{array}
\end{equation}
Compared to the CKMfitter result 
\begin{equation}
 \left|V_{td}/V_{ts}\right| = 0.2186^{+0.0049}_{-0.0059} \hspace{1cm} [\text{CKMfitter, tree}] \, ,
\label{eq:Vtq_ratio_tree}
\end{equation}
from the fit to tree-level processes our value \eqref{eq:VtdVts_result} is smaller by 
about 2.3 standard deviations. Thus, an improved determination of $|V_{td}|$ and $|V_{td}/V_{ts}|$ 
from tree-level processes might provide an interesting hint towards new physics in the $B_d$ system. 
Similar considerations have recently led to claims about an emerging $\Delta M_d$ anomaly 
\cite{Blanke:2018cya}. 

\begin{figure}[t]
 \begin{center}
    \includegraphics[width=0.7\textwidth]{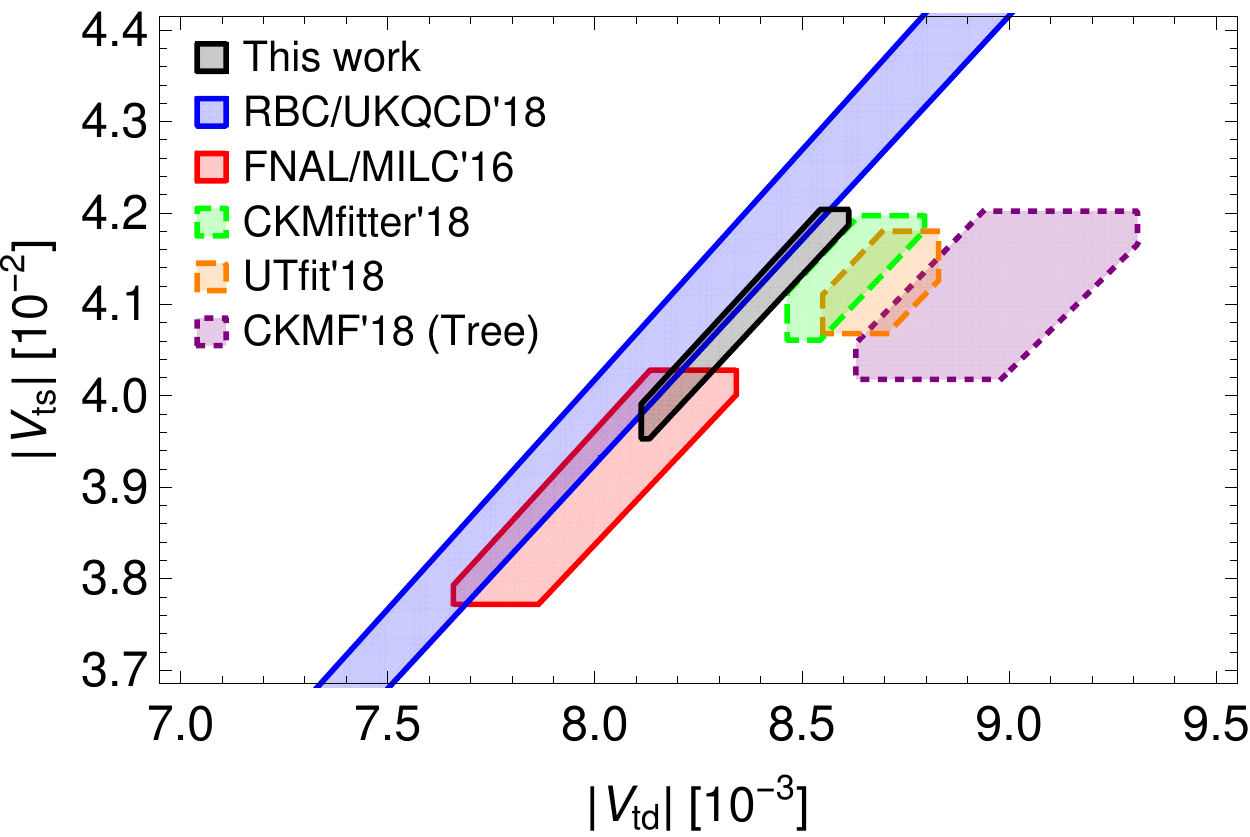}
  \caption{
  Comparison of our constraints on the CKM parameters $|V_{td}|$ and $|V_{ts}|$ with other works 
based on $B$ mixing~\cite{Bazavov:2016nty,Boyle:2018knm} (solid boundaries) and 
unitarity~\cite{Charles:2004jd,Bona:2006ah} (dashed boundaries). Since the full CKM fit includes 
the mass differences we also show the tree-level fit from CKMfitter~\cite{Charles:2004jd} (dotted 
boundaries). }
  \label{fig:Vts_Vtq_plane}
 \end{center}
\end{figure}

An overview of the various results is presented in Figure~\ref{fig:Vts_Vtq_plane}, where the 
overlap of the one-sigma regions for $|V_{td}|$, $|V_{ts}|$ and $|V_{td}/V_{ts}|$ is indicated 
by the shaded regions. Our results provide an important input for future CKM unitarity fits 
and can be used to extract the angle $\gamma$ in the unitarity triangle from the linear 
dependency between $\xi$ and the CKM angle $\gamma$ observed in \cite{Blanke:2016bhf}.

%


\subsection{\boldmath Determination of the top-quark $\overline{\text{MS}}$ mass\label{sec:mt}}

The parametric error from the top-quark mass currently dominates the uncertainty in the 
determination of the stability or meta-stability of the electroweak vacuum~\cite{Bednyakov:2015sca}. 
Direct measurements quote very precise values $m_t^\text{MC}=(173.0\pm0.4)$\,GeV for the top quark 
mass~\cite{Tanabashi:2018oca}, but these results correspond to so-called Monte-Carlo (MC) 
masses and not the top-quark pole mass. One therefore needs to account for additional uncertainties 
from the scheme conversion~\cite{Hoang:2018zrp} when these values are used for phenomenological 
predictions. Alternatively one can determine the top-quark mass by fitting observables like the 
total top-pair production cross section which can be predicted in terms of the top-quark mass 
in a well-defined scheme like $\overline{\text{MS}}$. Similarly, we can use the mass differences 
$\Delta M_q$ for a theoretically clean determination of $\overline{m}_t(\overline{m}_t)$. 
Using the CKMfitter values for $V_{td}$ and $V_{ts}$ as input we obtain 
\begin{equation} 
\begin{array}{ll}
 \overline{m}_t(\overline{m}_t) = (158_{-6}^{+9})\,\text{GeV} = (158_{-6}^{+7}\,(\text{had.})_{-1}^{+0}\,(\mu)_{-1}^{+6}\,(\text{param.}))\,\text{GeV}, \qquad & \text{from }\Delta M_s,\\
 \overline{m}_t(\overline{m}_t) = (155_{-6}^{+9})\,\text{GeV} = (155_{-6}^{+6}\,(\text{had.})_{-1}^{+0}\,(\mu)_{-2}^{+6}\,(\text{param.}))\,\text{GeV}, \qquad & \text{from }\Delta M_d.
\end{array}
\end{equation}
Combining both results we find 
\begin{equation}
\overline{m}_t(\overline{m}_t) = (157_{-6}^{+8})\,\text{GeV} = (157_{-6}^{+7}\,(\text{had.})_{-1}^{+0}\,(\mu)_{-1}^{+4}\,(\text{param.}))\,\text{GeV}, 
\end{equation}
where we have averaged over the hadronic and scale uncertainties, which are correlated, and 
treated the parametric uncertainties, which are dominated either by $V_{td}$ or $V_{ts}$, as 
independent. This is in good agreement with the PDG average~\cite{Tanabashi:2018oca}
\begin{equation}
 \overline{m}_t(\overline{m}_t) = (160_{-4}^{+5})\,\text{GeV}, 
\end{equation}
of $\overline{\text{MS}}$ mass determinations from cross section measurements with our 
uncertainty being about 50\% larger. A very precise measurement of the top-quark PS or 
$\overline{\text{MS}}$ mass with a total uncertainty of about 50 MeV is possible at a future 
lepton collider running at the top threshold~\cite{Beneke:2015kwa,Beneke:2017rdn,Simon:2016pwp}.


\subsection{\boldmath $\mathcal{B}(B_{q}\to\mu^+\mu^-)$\label{sec:bmumu}}

\begin{figure}[t]
 \begin{center}
    \includegraphics[width=0.7\textwidth]{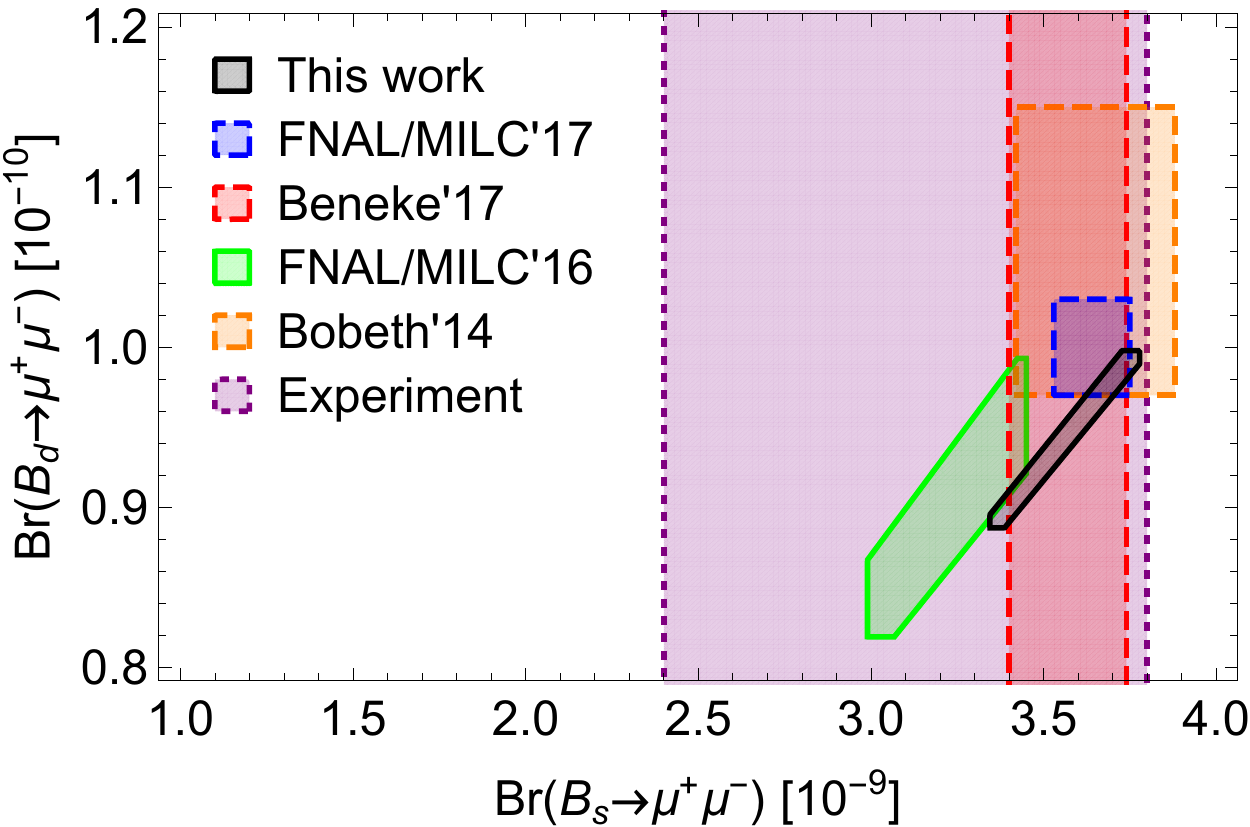}
  \caption{We compare our prediction for the branching ratios $\text{Br}(B_q^0\to \mu^+\mu^-)$ with $q=s,d$ 
  to other predictions using either the decay constants \cite{Bobeth:2013uxa,Beneke:2017vpq,Bazavov:2017lyh} 
  (dashed boundaries) or the Bag parameter $\overline{B}_{Q_1}^q$ \cite{Bazavov:2016nty} (solid boundaries) 
  as input. The experimental average for $\text{Br}(B_s^0\to \mu^+\mu^-)$ is indicated by the region with the 
  dotted boundary.}
  \label{fig:Bqll}
 \end{center}
\end{figure}

The branching ratio $\text{Br}(B_q\to l^+l^-)$ is strongly suppressed in the SM and theoretically 
clean. Thus, it provides a very sensitive probe for new physics. At present it has been computed 
at NNLO QCD plus NLO EW~\cite{Bobeth:2013uxa} and the dominant uncertainties are parametric, 
stemming from the decay constant and the CKM parameters. Both uncertainties cancel out 
of the ratio \cite{Buras:2003td}
\begin{equation}
 \frac{\text{Br}(B_q\to l^+l^-)}{\Delta M_q} = \frac{3G_F^2M_W^2m_l^2\tau_{B_q^H}}{\pi^3}\,\sqrt{1-\frac{4m_l^2}{M_{B_q}^2}}\,
                                                \frac{|C_A(\mu)|^2}{S_0(x_t)\hat{\eta}_B\overline{B}_{Q_1}^q(\mu)}\,,
\label{eq:BqToLL_Ratio}
\end{equation}
which in turn receives its dominant uncertainty from the Bag parameter $\overline{B}_{Q_1}^q$. 
Using our result \eqref{eq:DelB2_QCD_results_ratio} and including the power-enhanced QED 
corrections determined in \cite{Beneke:2017vpq} we predict the branching ratio by multiplying 
\eqref{eq:BqToLL_Ratio} with the measured mass differences  
\begin{align}
 \text{Br}(B_s^0\to \mu^+\mu^-)_\text{SM}  & = (3.55_{-0.20}^{+0.23})\cdot10^{-9}\,,\nonumber\\
 \text{Br}(B_d^0\to \mu^+\mu^-)_\text{SM}  & = (9.40_{-0.53}^{+0.58})\cdot10^{-11}\,,\nonumber\\
 \left(\frac{\text{Br}(B_d^0\to \mu^+\mu^-)}{\text{Br}(B_s^0\to \mu^+\mu^-)}\right)_\text{SM}  & = 0.0265\pm0.0003 
                                             = 0.0265\pm0.0002\,\left(\overline{B}_{Q_1}^{s/d}\right)\pm0.0002(\text{exp}) \,,
\label{eq:BqToMuMu_SM}
\end{align}
where the uncertainties for the branching ratios are completely dominated by the error from 
$\overline{B}_{Q_1}^q$. The result for $B_s^0\to \mu^+\mu^-$ is in good agreement with the current 
experimental average \cite{Amhis:2016xyh} 
\begin{equation}
 \text{Br}(B_s^0\to \mu^+\mu^-)_\text{exp} = (3.1\pm0.7)\cdot10^{-9}\,, 
\label{eq:BqToMuMu_exp}
\end{equation}
while the latest measurements only provide upper bounds at 95\% confidence level for $B_d^0\to \mu^+\mu^-$ 
\begin{equation}
\text{Br}(B_d^0\to \mu^+\mu^-)_\text{exp} < \begin{cases}
                                              11\cdot10^{-10}\,,\qquad\,\,(\text{CMS~\cite{Chatrchyan:2013bka}})\,,\\
                                             3.4\cdot10^{-10}\,,\qquad(\text{LHCb~\cite{Aaij:2017vad}})\,,\\
                                             2.1\cdot10^{-10}\,,\qquad(\text{ATLAS~\cite{Aaboud:2018mst}})\,.
                                            \end{cases}
\end{equation}
We compare our prediction \eqref{eq:BqToMuMu_SM} to the direct predictions from 
\cite{Bobeth:2013uxa,Beneke:2017vpq,Bazavov:2017lyh} which depend on the decay constants and CKM elements 
$|V_{tq}|$, the prediction \cite{Bazavov:2016nty} from the ratios $\text{Br}(B_q\to l^+l^-)/\Delta M_q$ 
and the experimental average \eqref{eq:BqToMuMu_exp} in Figure~\ref{fig:Bqll}. The shaded regions correspond 
to the overlap of the one-sigma regions for $\text{Br}(B_s^0\to \mu^+\mu^-)$, $\text{Br}(B_d^0\to \mu^+\mu^-)$ 
and $\text{Br}(B_d^0\to \mu^+\mu^-)/\text{Br}(B_s^0\to \mu^+\mu^-)$ where they were provided. We find good 
consistency among the various predictions with similar uncertainties for both approaches and good agreement 
with experiment whose uncertainty currently exceeds the theoretical one by a factor of about 3-4 in 
$\text{Br}(B_s^0\to \mu^+\mu^-)$.

For completeness we provide our predictions for the branching ratios to electrons 
\begin{align}
 \text{Br}(B_s^0\to e^+e^-)_\text{SM}  & = (8.37_{-0.48}^{+0.55})\cdot10^{-14}\,,\\
 \text{Br}(B_d^0\to e^+e^-)_\text{SM}  & = (2.22_{-0.13}^{+0.14})\cdot10^{-15}\,,\nonumber\\
 \left(\frac{\text{Br}(B_d^0\to e^+e^-)}{\text{Br}(B_s^0\to e^+e^-)}\right)_\text{SM}  & = 0.0265\pm0.0003 
                                             = 0.0265\pm0.0002\,\left(\overline{B}_{Q_1}^{s/d}\right)\pm0.0002(\text{exp}) \,,\nonumber
\label{eq:BqToEE_SM}
\end{align}
and tau leptons 
\begin{align}
 \text{Br}(B_s^0\to \tau^+\tau^-)_\text{SM}  & = (7.58_{-0.44}^{+0.50})\cdot10^{-7}\,,\\
 \text{Br}(B_d^0\to \tau^+\tau^-)_\text{SM}  & = (1.98_{-0.11}^{+0.12})\cdot10^{-8}\,,\nonumber\\
 \left(\frac{\text{Br}(B_d^0\to \tau^+\tau^-)}{\text{Br}(B_s^0\to \tau^+\tau^-)}\right)_\text{SM}  & = 0.0262\pm0.0003 
                                             = 0.0262\pm0.0002\,\left(\overline{B}_{Q_1}^{s/d}\right)\pm0.0002(\text{exp}) \,.\nonumber
\label{eq:BqToTauTau_SM}
\end{align}


\section{Conclusions\label{sec:conclusion}}
We have presented in this paper a HQET sum rule determination of
the five $\Delta B = 2$ Bag parameters describing $B_s$-mixing in the 
SM and beyond. For that we had to determine $m_s$ and $m_s^2$
corrections to the three-point correlator at the 3-loop level.
In particular we obtain the most precise values for the ratios
of Bag parameters in the $B_s$ and  $B_d$ system. Combing this 
result with the most recent lattice results for $f_{B_s}/f_{B_d}$
\cite{Aoki:2019cca,Hughes:2017spc,Bazavov:2017lyh} 
we obtain the world's most precise value for the ratio 
\begin{equation}
 \xi \equiv \frac{f_{B_s}}{f_B}\,\sqrt{\overline{B}_{Q_1}^{s/d}} = 1.2014_{-0.0072}^{+0.0065} \,,
\end{equation}
which represents a reduction of the uncertainty  by more than a factor of two 
compared to the latest lattice results \cite{Bazavov:2016nty,Boyle:2018knm}. 
Our results enable a rich phenomenology:
      we get updated SM predictions for the mixing observables
      $\Delta M_s$ and $\Delta \Gamma_s$, which are in agreement
      with the 
      experimental values. In particular we do not confirm the 
      large values for $\Delta M_s$ obtained with the 
      non-perturbative values from FNAL/MILC
      \cite{Bazavov:2016nty}, which led to severe bounds on 
      BSM models. If $V_{tb}$ and $\Delta M_q$ are used as
      inputs, we can precisely determine the CKM elements
      $|V_{td}|$ and $|V_{ts}|$ and we obtain the world's most
      precise determination of the ratio $|V_{td}/V_{ts}|$. 
      Using all CKM elements as inputs
      we get constraints on the values of the top quark $\overline{\text{MS}}$ 
      mass which are compatible with direct collider determinations.
      Finally our results lead also to precise SM predictions 
      for the branching ratios of the rare decays $B_q \to ll$.
      \\
      In future a still higher precision of our HEQT sum rule
      results can be obtained by the calculation of the HQET-QCD
      matching at NNLO (first steps in that direction have been
      performed in \cite{Grozin:2018wtg}). Another line of
      improvement could be the determination of
      $1/m_b$-corrections to the HQET limit. The computation of 
      $m_s$ corrections to the Bag parameters of $\Delta F=0$ 
      four-quark operators would enable an update of the predictions 
      for the lifetime ratios $\tau(B_s)/\tau(B^0)$ \cite{Kirk:2017juj} 
      and $\tau(D_s^+)/\tau(D^0)$ \cite{Lenz:2013aua}. 
      Finally a cross-check of our HQET sum results for 
      mixing and lifetimes with modern lattice techniques 
      would be very desirable.

\subsection*{Acknowledgements}

We thank S\'ebastien Descotes-Genon for providing us with the result \eqref{eq:Vtq_ratio_tree}, 
Oliver Witzel for interesting discussions and Andrzej Buras, Aleksei Rusov and Tobias Tsang for comments on 
the manuscript. This work was supported by the STFC through the IPPP grant and a Postgraduate Studentship.

\newpage
\appendix


\section{Inputs and detailed overview of uncertainties\label{sec:Uncertainties}}

\begin{longtable}{|c|cc|}
\hline  
\rule{0pt}{3ex} Parameter                               & \hspace{2cm}Value\hspace{2cm}       & Source \\ \hline
\rule{0pt}{3ex} $\overline{m}_b(\overline{m}_b)$        & $(4.203_{-0.034}^{+0.016})$ GeV     & \cite{Beneke:2014pta,Beneke:2016oox} \\ 
\rule{0pt}{3ex} $m_b^\text{PS}(2\text{ GeV})$           & $(4.532_{-0.039}^{+0.013})$ GeV     & \cite{Beneke:2014pta,Beneke:2016oox} \\ 
\rule{0pt}{3ex} $\overline{m}_c(\overline{m}_c)$        & $(1.279\pm0.013)$ GeV               & \cite{Chetyrkin:2009fv} \\
\rule{0pt}{3ex} $m_t^\text{pole}$                       & $(173.0\pm0.4)$ GeV                 & \cite{Tanabashi:2018oca} \\ 
\rule{0pt}{3ex} $\alpha_s(M_Z)$                         & $0.1181\pm0.0011$                   & \cite{Tanabashi:2018oca} \\ 
\rule{0pt}{3ex} $V_{us}$                                & $0.224745^{+0.000254}_{-0.000059}$  & \cite{Charles:2004jd} \\ 
\rule{0pt}{3ex} $V_{ub}$                                & $0.003746^{+0.000090}_{-0.000062}$  & \cite{Charles:2004jd} \\ 
\rule{0pt}{3ex} $V_{cb}$                                & $0.04240^{+0.00030}_{-0.00115}$     & \cite{Charles:2004jd} \\ 
\rule{0pt}{3ex} $\gamma$                                & $(65.81^{+0.99}_{-1.66})^\circ$     & \cite{Charles:2004jd} \\ 
\rule{0pt}{3ex} $f_{B}$                                 & $(190.0\pm1.3)$ MeV                 & \cite{Aoki:2019cca} \\ 
\rule{0pt}{3ex} $f_{B_s}$                               & $(230.3\pm1.3)$ MeV                 & \cite{Aoki:2019cca} \\ 
\rule{0pt}{3ex} $f_{B_s}/f_B$                           & $1.209\pm0.005$                     & \cite{Aoki:2019cca} \\ 
\rule{0pt}{3ex} $\tau(B_s^{0,\text{H}})$                & $(1.615\pm0.009)\text{ ps}^{-1}$    & \cite{Tanabashi:2018oca} \\ 
\rule{0pt}{3ex} $\tau(B_d^0)$                           & $(1.520\pm0.004)\text{ ps}^{-1}$    & \cite{Tanabashi:2018oca} \\ 
\hline
\caption{Input values for parameters.\label{tab:inputs}}
\end{longtable}

\begin{longtable}[h!]{|l|cccccccc|}
\hline \rule{0pt}{3ex}                 & $\Lbar$              & intrinsic SR & condensates & $\mu_\rho$           & $m_s$                & $1/m_b$    & $\mu_m$              & $a_i$                \\ \hline
\rule{0pt}{3ex} $\overline{B}_{Q_1}^s$ & $_{-0.003}^{+0.002}$ & $\pm0.018$   & $\pm0.004$  & $_{-0.027}^{+0.013}$ & $_{-0.002}^{+0.003}$ & $\pm0.010$ & $_{-0.038}^{+0.044}$ & $_{-0.008}^{+0.007}$ \\ 
\rule{0pt}{3ex} $\overline{B}_{Q_2}^s$ & $_{-0.014}^{+0.012}$ & $\pm0.020$   & $\pm0.004$  & $_{-0.015}^{+0.010}$ & $_{-0.004}^{+0.004}$ & $\pm0.010$ & $_{-0.063}^{+0.072}$ & $_{-0.015}^{+0.015}$ \\ 
\rule{0pt}{3ex} $\overline{B}_{Q_3}^s$ & $_{-0.055}^{+0.047}$ & $\pm0.107$   & $\pm0.023$  & $_{-0.001}^{+0.026}$ & $_{-0.026}^{+0.024}$ & $\pm0.010$ & $_{-0.073}^{+0.091}$ & $_{-0.053}^{+0.054}$ \\ 
\rule{0pt}{3ex} $\overline{B}_{Q_4}^s$ & $_{-0.005}^{+0.006}$ & $\pm0.021$   & $\pm0.011$  & $_{-0.002}^{+0.000}$ & $_{-0.002}^{+0.003}$ & $\pm0.010$ & $_{-0.079}^{+0.088}$ & $_{-0.006}^{+0.006}$ \\ 
\rule{0pt}{3ex} $\overline{B}_{Q_5}^s$ & $_{-0.012}^{+0.014}$ & $\pm0.018$   & $\pm0.009$  & $_{-0.007}^{+0.000}$ & $_{-0.006}^{+0.007}$ & $\pm0.010$ & $_{-0.067}^{+0.075}$ & $_{-0.012}^{+0.012}$ \\
\hline
\caption{Individual errors for the Bag parameters in the $B_s$ system.\label{tab:details_Bag}}
\end{longtable}

\begin{longtable}[h!]{|l|cccccccc|}
\hline \rule{0pt}{3ex}                     & $\Lbar$              & intrinsic SR & condensates & $\mu_\rho$           & $m_s$                & $1/m_b$    & $\mu_m$              & $a_i$                \\ \hline
\rule{0pt}{3ex} $\overline{B}_{Q_1}^{s/d}$ & $_{-0.002}^{+0.001}$ & $\pm0.005$   & $\pm0.002$  & $_{-0.006}^{+0.002}$ & $_{-0.002}^{+0.003}$ & $\pm0.002$ & $_{-0.000}^{+0.000}$ & $_{-0.000}^{+0.000}$ \\ 
\rule{0pt}{3ex} $\overline{B}_{Q_2}^{s/d}$ & $_{-0.003}^{+0.004}$ & $\pm0.005$   & $\pm0.002$  & $_{-0.002}^{+0.005}$ & $_{-0.004}^{+0.005}$ & $\pm0.002$ & $_{-0.000}^{+0.000}$ & $_{-0.000}^{+0.000}$ \\ 
\rule{0pt}{3ex} $\overline{B}_{Q_3}^{s/d}$ & $_{-0.023}^{+0.036}$ & $\pm0.025$   & $\pm0.010$  & $_{-0.019}^{+0.042}$ & $_{-0.031}^{+0.029}$ & $\pm0.002$ & $_{-0.005}^{+0.004}$ & $_{-0.005}^{+0.005}$ \\ 
\rule{0pt}{3ex} $\overline{B}_{Q_4}^{s/d}$ & $_{-0.002}^{+0.001}$ & $\pm0.005$   & $\pm0.002$  & $_{-0.005}^{+0.002}$ & $_{-0.002}^{+0.003}$ & $\pm0.002$ & $_{-0.000}^{+0.000}$ & $_{-0.000}^{+0.000}$ \\ 
\rule{0pt}{3ex} $\overline{B}_{Q_5}^{s/d}$ & $_{-0.004}^{+0.003}$ & $\pm0.005$   & $\pm0.002$  & $_{-0.010}^{+0.004}$ & $_{-0.006}^{+0.006}$ & $\pm0.002$ & $_{-0.000}^{+0.000}$ & $_{-0.000}^{+0.000}$ \\
\hline
\caption{Individual errors for the ratio of Bag parameters in the $B_s$ and $B_d$ system.\label{tab:details_Bag_ratio}}
\end{longtable}

\begin{longtable}[h!]{|l|cccc|}
\hline \rule{0pt}{3ex}                    & $\Delta M_s^\text{SM}$ [ps$^{-1}$] & $\Delta \Gamma_s^\text{PS}$ [ps$^{-1}$] & $\Delta M_d^\text{SM}$ [ps$^{-1}$] & $\Delta \Gamma_d^\text{SM}$ [$10^{-3}$ps$^{-1}$] \\ \hline
\rule{0pt}{3ex} $\overline{B}_{Q_1}^q$    & $\pm1.1$                           & $\pm0.005$                              & $\pm0.031$                         & $_{-0.15}^{+0.16}$ \\ 
\rule{0pt}{3ex} $\overline{B}_{Q_3}^q$    & $\pm0.0$                           & $_{-0.005}^{+0.006}$                    & $\pm0.000$                         & $_{-0.16}^{+0.17}$ \\ 
\rule{0pt}{3ex} $\overline{B}_{R_0}^q$    & $\pm0.0$                           & $\pm0.004$                              & $\pm0.000$                         & $\pm0.10$ \\ 
\rule{0pt}{3ex} $\overline{B}_{R_1}^q$    & $\pm0.0$                           & $\pm0.000$                              & $\pm0.000$                         & $\pm0.01$ \\
\rule{0pt}{3ex} $\overline{B}_{R_1'}^q$   & $\pm0.0$                           & $\pm0.000$                              & $\pm0.000$                         & $\pm0.01$ \\
\rule{0pt}{3ex} $\overline{B}_{R_2}^q$    & $\pm0.0$                           & $\pm0.018$                              & $\pm0.000$                         & $\pm0.53$ \\
\rule{0pt}{3ex} $\overline{B}_{R_3}^q$    & $\pm0.0$                           & $\pm0.000$                              & $\pm0.000$                         & $\pm0.00$ \\
\rule{0pt}{3ex} $\overline{B}_{R_3'}^q$   & $\pm0.0$                           & $\pm0.000$                              & $\pm0.000$                         & $\pm0.01$ \\
\rule{0pt}{3ex} $f_{B_q}$                 & $\pm0.2$                           & $\pm0.001$                              & $_{-0.007}^{+0.008}$               & $\pm0.04$ \\ 
\rule{0pt}{3ex} $\mu_1$                   & $\pm0.0$                           & $_{-0.021}^{+0.008}$                    & $\pm0.000$                         & $_{-0.60}^{+0.24}$ \\ 
\rule{0pt}{3ex} $\mu_2$                   & $\pm0.1$                           & $_{-0.003}^{+0.000}$                    & $_{-0.002}^{+0.004}$               & $_{-0.08}^{+0.00}$ \\ 
\rule{0pt}{3ex} $m_b$                     & $\pm0.0$                           & $_{-0.001}^{+0.000}$                    & $\pm0.000$                         & $_{-0.04}^{+0.01}$ \\ 
\rule{0pt}{3ex} $m_c$                     & $\pm0.0$                           & $\pm0.001$                              & $\pm0.000$                         & $\pm0.02$ \\ 
\rule{0pt}{3ex} $\alpha_s$                & $\pm0.0$                           & $\pm0.000$                              & $\pm0.001$                         & $\pm0.01$ \\ 
\rule{0pt}{3ex} CKM                       & $_{-1.0}^{+0.3}$                   & $_{-0.005}^{+0.001}$                    & $_{-0.032}^{+0.011}$               & $_{-0.15}^{+0.06}$ \\ 
\hline
\caption{Individual errors for the $B_s$ and $B_d$ mixing observables.\label{tab:details_mixing}}
\end{longtable}

\end{document}